\documentclass{article}
\usepackage{epubtk}
\usepackage{booktabs}
\usepackage{graphicx}
\usepackage{textcomp}

\newcommand{\kmsmpc}{\ensuremath{\mathrm{\ km\ s}^{-1}\mathrm{\ Mpc}^{-1}}}

\begin{document}

\title{The Hubble Constant}

\author{\epubtkAuthorData{Neal Jackson}
       {Jodrell Bank Centre for Astrophysics \\
        School of Physics and Astronomy\\
        University of Manchester \\
        Turing Building \\
        Manchester M13 9PL, U.K.}
       {njj@jb.man.ac.uk}
       {http://www.jodrellbank.manchester.ac.uk/~njj/}}

\date{}

\maketitle


\begin{abstract}
I review the current state of determinations of the Hubble constant,
which gives the length scale of the Universe by relating the
expansion velocity of objects to their distance. There are two broad
categories of measurements. The first uses individual astrophysical objects which
have some property that allows their intrinsic luminosity or size to be
determined, or allows the determination of their distance by geometric 
means. The second category comprises the use of all-sky cosmic microwave
background, or correlations between large samples of galaxies, to determine
information about the geometry of the Universe and hence the Hubble constant,
typically in a combination with other cosmological parameters. Many, but not
all, object-based measurements give $H_0$ values of around 72-74\kmsmpc\ , with
typical errors of 2-3\kmsmpc. This is in mild discrepancy with CMB-based measurements, 
in particular those from the Planck satellite, which give values of 67-68\kmsmpc\ and
typical errors of 1-2\kmsmpc. The size of the remaining systematics indicate that
accuracy rather than precision is the remaining problem in a good determination
of the Hubble constant. Whether a discrepancy exists, and whether new physics
is needed to resolve it, depends on details of the systematics of the object-based 
methods, and also on the assumptions about
other cosmological parameters and which datasets are combined in the case of
the all-sky methods.
\end{abstract}

\epubtkKeywords{Cosmology, Hubble constant}

\newpage


\section{Introduction}


\subsection{A brief history}

The last century saw an expansion in our view of the world from a
static, Galaxy-sized Universe, whose constituents were stars and 
``nebulae'' of unknown but possibly stellar origin, to the view 
that the observable Universe is in a state of expansion from an initial
singularity over ten billion years ago, and contains approximately
100~billion galaxies. This paradigm shift was summarised in a famous
debate between Shapley and Curtis in 1920; summaries of the views of
each protagonist can be found in~\cite{curti20} and~\cite{shapl19}.

The historical background to this change in world view has been
extensively discussed and whole books have been devoted to the subject
of distance measurement in astronomy~\cite{rowan85}. At the heart 
of the change was the conclusive proof that what we now know as external
galaxies lay at huge distances, much greater than those between
objects in our own Galaxy. The earliest such distance determinations
included those of the galaxies NGC~6822~\cite{hubbl25},
M33~\cite{hubbl26} and M31~\cite{hubbl29a}, by Edwin Hubble.

As well as determining distances, Hubble also considered redshifts of
spectral lines in galaxy spectra which had previously been measured by
Slipher in a series of papers~\cite{sliph14,sliph17}. If a
spectral line of emitted wavelength $\lambda_0$ is observed at a
wavelength $\lambda$, the redshift $z$ is defined as
\begin{equation}
z = \lambda/\lambda_0 - 1.
\end{equation}
For nearby objects and assuming constant gravitational tidal field, the redshift 
may be thought of as corresponding to a recession velocity $v$ which for nearby 
objects behaves in a way predicted by a simple Doppler formula\epubtkFootnote{See 
\cite{kaise14} and references therein, e.g. \cite{bunn09}, for discussion of the 
details of the interpretation of redshift in an expanding Universe. The first level 
of sophistication involves maintaining the GR principle that space is locally 
Minkowskian, so that in a small region of space all effects must reduce to SR (for 
instance, in Peacock's \cite{peaco99} 
example, the expansion of the Universe does not imply that a long-lived human will grow 
to four metres tall in the next $10^{10}$~years). Redshift can be thought of as a series 
of transformations in photon wavelengths between an infinite succession of closely-separated 
observers, resulting in an overall wavelength shift between two observers with a finite
separation and therefore an associated ``velocity''. The second level 
of sophistication is to ask what this velocity actually represents. \cite{kaise14} 
calculates the ratio of photon wavelength shifts between pairs of fundamental observers
to the shifts in their proper separation in the presence of arbitrary gravitational 
fields, and shows that this ratio only corresponds to the purely dynamical result if
the gravitational tide is constant.}, $v=cz$. Hubble showed that a relation existed 
between distance and redshift (see Figure~\ref{figure_01}); more distant galaxies 
recede faster, an observation which can naturally be explained if the Universe as a 
whole is expanding. The relation between the recession velocity and distance is linear 
in nearby objects, as it must be if the same dependence is to be observed from any other 
galaxy as it is from our own Galaxy (see Figure~\ref{figure_02}).
The proportionality constant is the Hubble constant $H_0$, where the 
subscript indicates a value as measured now. Unless the Universe's 
expansion does not accelerate or decelerate, the slope of the 
velocity--distance relation is different for observers at different 
epochs of the Universe. As well as the velocity corresponding to the universal expansion, a galaxy
also has a ``peculiar velocity'', typically of a few hundred kms\super{-1}, due to groups or
clusters of galaxies in its vicinity. Peculiar velocities are a nuisance if determining the
Hubble constant from relatively nearby objects for which they are comparable to the recession
velocity. Once the distance is $>$~50~Mpc, the recession velocity is large enough for the
error in $H_0$ due to the peculiar velocity to be less than about 10\%.

\epubtkImage{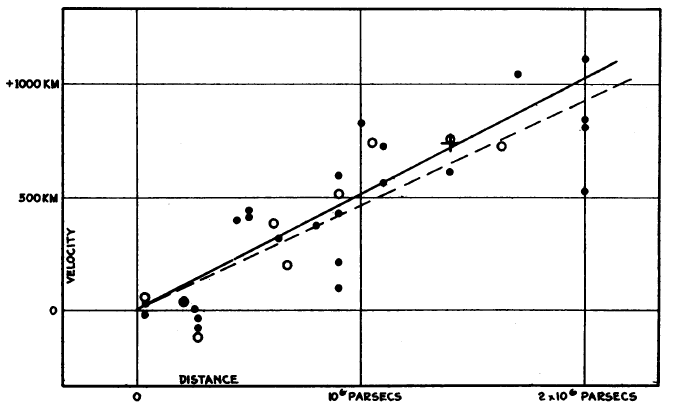}{%
  \begin{figure}[htbp]
    \centerline{\includegraphics[scale=0.5]{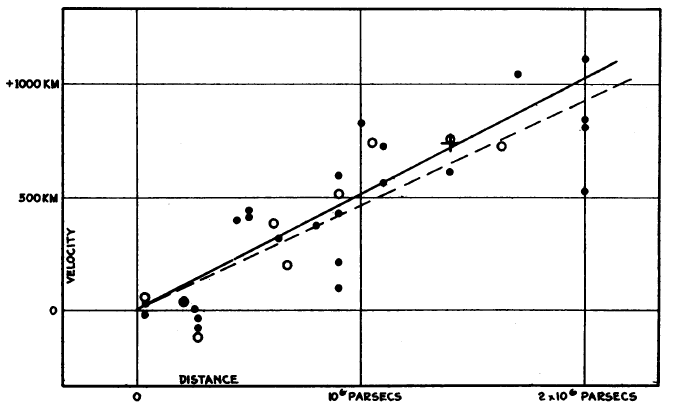}}
    \caption{Hubble's original diagram of distance to nearby
      galaxies, derived from measurements using Cepheid variables,
      against velocity, derived from redshift~\cite{hubbl29b}. The
      Hubble constant is the slope of this relation, and in this
      diagram is a factor of nearly 10 steeper than currently accepted
      values.}
    \label{figure_01}
\end{figure}}

\epubtkImage{h0_hlaw.png}{%
  \begin{figure}[htbp]
    \centerline{\includegraphics[scale=0.6]{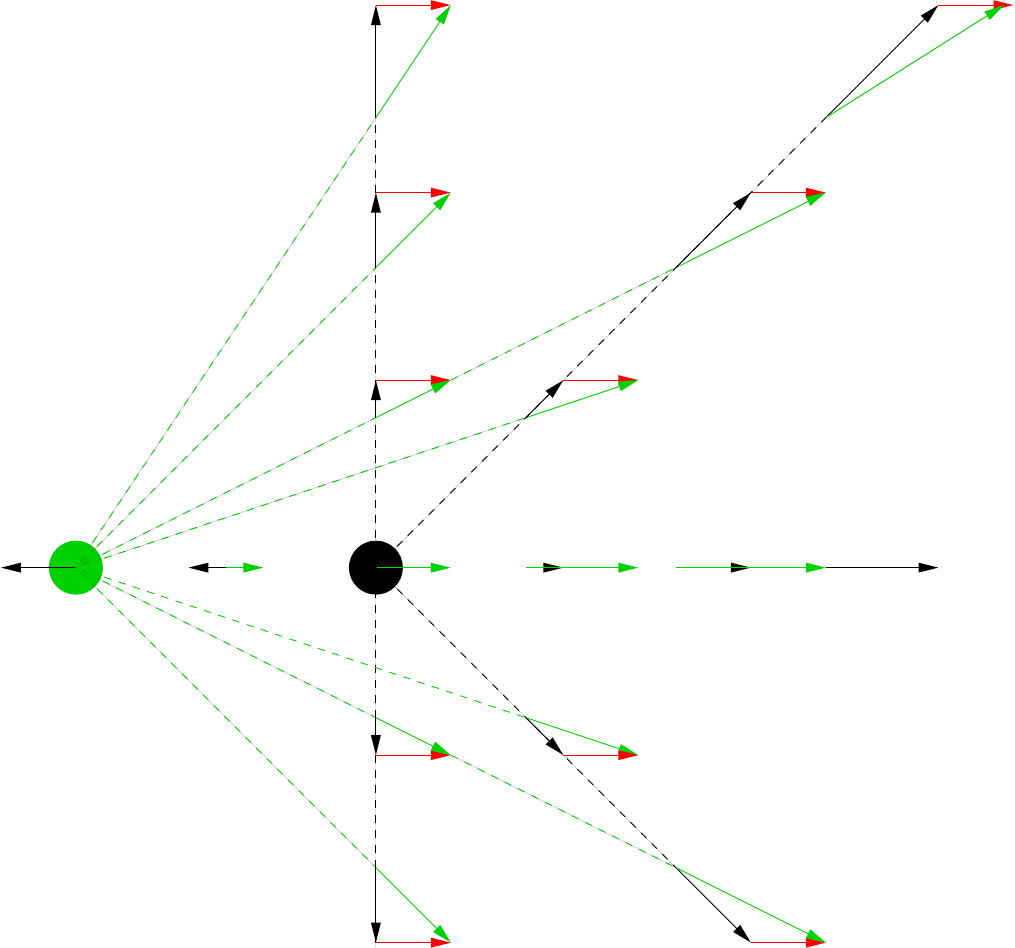}}
    \caption{Illustration of the Hubble law. Galaxies at all
      points of the square grid are receding from the black galaxy at
      the centre, with velocities proportional to their distance away
      from it. From the point of view of the second, green, galaxy two
      grid points to the left, all velocities are modified by vector
      addition of its velocity relative to the black galaxy (red
      arrows). When this is done, velocities of galaxies as seen by
      the second galaxy are indicated by green arrows; they all appear
      to recede from this galaxy, again with a Hubble-law linear
      dependence of velocity on distance.}
    \label{figure_02}
\end{figure}}


Recession velocities are very easy to measure; all we need is an object 
with an emission line and a spectrograph. Distances are very difficult.
This is because in order to measure a distance, we need a standard
candle (an object whose luminosity is known) or a standard ruler (an
object whose length is known), and we then use apparent brightness or
angular size to work out the distance. Good standard candles and standard
rulers are in short supply because most such objects require that we 
understand their astrophysics well enough to work out what their luminosity or 
size actually is. Neither stars nor galaxies by themselves remotely 
approach the uniformity needed; even when selected by other, easily
measurable properties 
such as colour, they range over orders of magnitude in luminosity and
size for reasons that are astrophysically interesting but frustrating 
for distance measurement. The ideal $H_0$ object, in fact, is one which 
involves as little astrophysics as possible.

Hubble originally used a class of stars known as Cepheid variables for
his distance determinations. These are giant blue stars, the best known
of which is $\alpha$UMa, or Polaris. In most normal stars, a
self-regulating mechanism exists in which any tendency for the star to
expand or contract is quickly damped out. In a small range of temperature
on the Hertzsprung--Russell (H-R) diagram, around 7000\,--\,8000~K, particularly at 
high luminosity,\epubtkFootnote{This is known in the literature as the
``instability strip'' and is almost, but not quite, parallel to the
luminosity axis on the H-R diagram. In normal stars, any compression of
the star, and the associated rise in temperature, results in a decrease
in opacity; the resulting escape of photons produces expansion and cooling.
For stars in the instability strip, a layer of partially ionized He close
to the surface causes opacity to rise instead of falling with an increase
in temperature, producing a degree of positive feedback and consequently
oscillations. The instability strip has a finite width, which
causes a small degree of dispersion in period--luminosity correlations
among Cepheids.} this does not happen and pulsations occur. These 
pulsations, the defining property of Cepheids, have a characteristic form,
a steep rise followed by a gradual fall. They also have a period which is
directly proportional to luminosity, because brighter stars are larger, 
and therefore take longer to pulsate. The period-luminosity relationship
was discovered by Leavitt~\cite{leavi12} by studying a sample of Cepheid
variables in the Large Magellanic Cloud (LMC). Because these stars were
known to be all at the same distance, their correlation of apparent
magnitude with period therefore implied the \textit{P-L}~relationship.

The Hubble constant was originally measured as
$500\kmsmpc$~\cite{hubbl29b}
and its subsequent history was a more-or-less uniform revision 
downwards. In the early days this was caused by bias\epubtkFootnote{There are
numerous subtle and less-subtle biases in distance measurement; 
see~\cite{teeri97} for a blow-by-blow account. The simplest bias, the
``classical'' Malmquist bias, arises because, in any population of objects with a 
distribution in intrinsic luminosity, only the brighter members of the 
population will be seen at large distances. The result is that the
inferred average luminosity is greater than the true luminosity,
biasing distance measurements towards the systematically short. The
Behr bias~\cite{behr51} from 1951 is a distance-dependent
version of the Malmquist bias, namely that at higher distances,
increasingly bright galaxies will be missing from samples.
This leads to an overestimate of the average brightness of the standard 
candle which becomes worse at higher distance.} 
in the original samples~\cite{behr51}, confusion between bright stars and 
H{\sc ii} regions in the original samples~\cite{humas56,sanda58} 
and differences between type~I and~II 
Cepheids\epubtkFootnote{Cepheids come in two flavours: type~I and type~II,
corresponding to population I and II stars. 
Population~II stars are an earlier metal-poor generation of stars,
which formed after the hypothetical, truly primordial Population~III stars,
but before later-generation Population~I stars like the Sun which contain 
significant extra amounts of elements other than hydrogen and
helium due to enrichment of the ISM by supernovae in the meantime. 
The name ``Cepheid'' derives from the fact that the star
$\delta$~Cephei was the first to be identified (by Goodricke in 1784).
Population~II Cepheids are sometimes known as W~Virginis stars,
after their prototype, W~Vir, and a W~Vir star is typically a factor 
of 3 fainter than a classical Cepheid of the same
period.} \cite{baade56}. In the second half of
the last century, the subject was dominated by a lengthy 
dispute between investigators favouring values around
$50\kmsmpc$ and those preferring
higher values of $100\kmsmpc$. Most
astronomers would now bet large amounts of money on the true value lying
between these extremes, and this review is an attempt to explain why and
also to try and evaluate the evidence for the best-guess 
current value. It is not an attempt to review the global history of $H_0$ 
determinations, as this has been done many times, often by the original 
protagonists or their close collaborators. For an overall review of this 
process see, for example, \cite{trimb96} and~\cite{tamma05}. 
Compilations of data and analysis of them are given by Huchra
(\url{http://cfa-www.harvard.edu/~huchra/hubble}),
Allen (\url{http://www.institute-of-brilliant-failures.com/}), and Gott (\cite{gott01}, updated
by~\cite{chen03}).\epubtkFootnote{The conclusion of the latter, that based on median statistics
of the Huchra compilation, $H_0=67 \pm 2\kmsmpc$ , is slightly scary in retrospect given
the Planck value of $67.2 \pm 1.3\kmsmpc$ for a flat Universe~\cite{planc13}.} Further reviews of the 
subject, with various different emphases and approaches, are given by 
\cite{tamma08,freed10}.

In summary, the ideal object for measuring the Hubble constant:

\begin{itemize}
\item Has a property which allows it to be treated as either as a standard candle or as a 
standard ruler
\item Can be used independently of other calibrations (i.e., in a one-step process)
\item Lies at a large enough distance (a few tens of Mpc or greater) that peculiar
velocities are small compared to the recession velocity at that distance
\item Involves as little astrophysics as possible, so that the distance determination
does not depend on internal properties of the object
\item Provides the Hubble constant independently of other cosmological parameters.
\end{itemize}

Many different methods are discussed in this review. We begin with one-step methods, and in 
particular with the use of megamasers in external galaxies -- arguably the only method 
which satisfies all the above criteria. Two other one-step methods, gravitational lensing 
and Sunyaev--Zel'dovich measurements, which have significant contaminating astrophysical 
effects are also discussed. The review then discusses two other programmes: first, the 
Cepheid-based distance ladders, where the astrophysics is probably now well understood after 
decades of effort, but which are not one-step processes; and second, information from the CMB, 
an era where astrophysics is in the linear regime and therefore simpler, but where $H_0$ is 
not determined independently of other 
cosmological parameters in a single experiment, without further assumptions.


\subsection{A little cosmology}

The expanding Universe is a consequence, although not the only possible
consequence, of general relativity coupled with the assumption that
space is homogeneous (that is, it has the same average density of matter
at all points at a given time) and isotropic (the same in all
directions). In 1922, Friedman~\cite{fried22} showed that given that assumption, we can
use the Einstein field equations of general relativity to write down the
dynamics of the Universe using the following two equations, now known as the
Friedman equations:
\begin{eqnarray}
  \dot{a}^2 - \frac{1}{3}(8 \pi G \rho + \Lambda) a^2 & = &
  - k c^2,
  \\
  \frac{\ddot{a}}{a} & = &
  - \frac{4}{3} \pi G (\rho + 3 p / c^2) + \frac{1}{3} \Lambda.
\end{eqnarray}%
Here $a=a(t)$ is the scale factor of the Universe. It is fundamentally 
related to redshift, because the quantity $(1+z)$ is the ratio of the 
scale of the Universe now to the scale of the Universe at the time of 
emission of the light ($a_0/a$). 
$\Lambda$ is the cosmological constant, which appears in the field 
equation of general relativity as an extra term. It corresponds to a
universal repulsion and was originally introduced by Einstein to coerce
the Universe into being static. On Hubble's discovery of the expansion 
of the Universe, he removed it, only for it to reappear seventy years 
later as a result of new data~\cite{perlm97,riess98} (see 
also~\cite{carro01,weinb13} for a review). $k$ is a curvature term, and
is $-1$, $0$, or $+1$, according to whether the global geometry of the 
Universe is negatively curved, spatially flat, or positively curved. 
$\rho$ is the density of the contents of the Universe, 
$p$ is the pressure and dots represent time derivatives. For any
particular component of the Universe, we need to specify an equation for 
the relation of pressure to density to solve these equations; for most 
components of interest such an equation is of the form $p=w\rho$.
Component densities vary with scale factor $a$ as the Universe expands,
and hence vary with time.

At any given time, we can define a Hubble parameter
\begin{equation}
  H(t) = \dot{a} / a,
\end{equation}
which is obviously related to the Hubble constant, because it is the
ratio of an increase in scale factor to the scale factor itself. In
fact, the Hubble constant $H_0$ is just the value of $H$ at the
current time\epubtkFootnote{Historically, the Hubble constant has often
been quoted as $H_0$=100$h$\kmsmpc, as a way of maintaining agnosticism
in an era where observations allowed a wide range in $h$. This is largely
disappearing, but papers using $h$ can be hard to interpret~\cite{croto13}}.

If $\Lambda=0$, we can derive the kinematics of the Universe quite
simply from the first Friedman equation. For a spatially flat Universe
$k=0$, and we therefore have
\begin{equation}
  \rho = \rho_{\mathrm{c}} \equiv \frac{3 H^2}{8 \pi G},
  \label{equation_05}
\end{equation}
where $\rho_{\mathrm{c}}$ is known as the critical density. For Universes whose
densities are less than this critical density, $k<0$ and space is negatively
curved. For such Universes it is easy to see from the first Friedman
equation that we require $\dot{a}>0$, and therefore the Universe
must carry on expanding for ever. For positively curved Universes
($k>0$), the right hand side is negative, and we reach a point at which
$\dot{a}=0$. At this point the expansion will stop and thereafter go
into reverse, leading eventually to a Big Crunch as $\dot{a}$ becomes
larger and more negative.

For the global history of the Universe in models with a cosmological
constant, however, we need to consider the $\Lambda$ term as providing
an effective acceleration. If the cosmological constant is
positive, the Universe is almost bound to expand forever, unless the
matter density is very much greater than the energy density in 
cosmological constant and can collapse the Universe before the
acceleration takes over. (A negative cosmological constant will always
cause recollapse, but is not part of any currently likely world model). 
Carroll~\cite{carro01} provides further discussion of this point.

We can also introduce some dimensionless symbols for energy densities in 
the cosmological constant at the current time,
$\Omega_{\Lambda}\equiv\Lambda/(3H_0^2)$, and in
``curvature energy'', $\Omega_k\equiv -kc^2/H_0^2$. By rearranging the
first Friedman equation we obtain
%
\begin{equation}
  \frac{H^2}{H_0^2} =
  \frac{\rho}{\rho_{\mathrm{c}}} - \Omega_k a^{-2} + \Omega_\Lambda.
  \label{equation_06}
\end{equation}

The density in a particular component of the Universe $X$, as a
fraction of critical density, can be written as
\begin{equation}
  \rho_X / \rho_{\mathrm{c}} = \Omega_X a^\alpha,
  \label{equation_07}
\end{equation}
where the exponent $\alpha$ represents the dilution of the component as 
the Universe expands. It is related to the $w$ parameter defined earlier 
by the equation $\alpha=-3(1+w)$. For ordinary matter $\alpha=-3$, and for
radiation $\alpha=-4$, because in addition to geometrical dilution as the
universe expands, the energy of radiation decreases as the wavelength 
increases. The cosmological constant 
energy density remains the same no matter how the size of the Universe 
increases, hence for a cosmological constant we have $\alpha=0$ and $w=-1$. 
$w=-1$ is not the only possibility for producing acceleration, however.
Any general class of ``quintessence'' models for which $w<-\frac{1}{3}$ will 
do; the case $w<-1$ is probably the most extreme and eventually results in the 
accelerating expansion becoming so dominant that all gravitational interactions become
impossible due to the shrinking boundary of the observable Universe, finally
resulting in all matter being torn apart in a ``Big Rip'' \cite{caldw03}.
In current models $\Lambda$ will become increasingly dominant in the
dynamics of the Universe as it expands. Note that
\begin{equation}
\sum_X \Omega_X + \Omega_\Lambda + \Omega_k = 1
\label{equation_07A}
\end{equation}
by definition, because $\Omega_k=0$ implies a flat Universe in
which the total energy density in matter together with the cosmological
constant is equal to the critical density. Universes for which
$\Omega_k$ is 
almost zero tend to evolve away from this point, so the observed
near-flatness is a puzzle known as the ``flatness problem''; the
hypothesis of a period of rapid expansion known as inflation in the 
early history of the Universe predicts this near-flatness naturally.
Recent work in which a detection of CMB B-modes with 
BICEP2 is claimed~\cite{ade14} supports (if confirmed) the inflation hypothesis.
As well as a solution to the flatness problem, inflation is an attractive idea because it
provides a natural explanation for the large-scale uniformity of the Universe
in regions which would otherwise not be in causal contact with each other.

We finally obtain an equation for the variation of the Hubble parameter
with time in terms of the Hubble constant (see e.g.~\cite{peaco99}),
\begin{equation}
  H^2 = H_0^2(\Omega_{\Lambda} + \Omega_\mathrm{m}a^{-3} +
  \Omega_\mathrm{r}a^{-4} + \Omega_ka^{-2}),
\label{equation_07B}
\end{equation}
%
where $\Omega_{\mathrm{r}}$ represents the energy density in
radiation and $\Omega_{\mathrm{m}}$ the energy density in matter.

To obtain cosmological distances, we need to perform integrals of the form

\begin{equation}
D_C = c\int{\frac{dz}{H(z)}}
\label{equation_07C}
\end{equation}

where the right-hand side can be expressed as a ``Hubble distance'' $D_H\equiv c/H_0$,
multiplied by an integral over dimensionless quantities such as the $\Omega$ 
terms. We can define a number of distances in cosmology, including the ``comoving''
distance $D_C$ defined above. The most important for
present purposes are the angular diameter distance $D_{\mathrm{A}}=D_{\mathrm C}/(1+z)$,
which relates the apparent angular size of an object to its proper
size, and the luminosity distance $D_{\mathrm{L}}=(1+z)^2D_{\mathrm{A}}$, 
which relates the observed flux of an object to its intrinsic luminosity.
For currently popular models, the angular diameter distance increases to
a maximum as $z$ increases to a value of order 1, and decreases thereafter.
Formulae for, and fuller explanations of, both distances are given
by~\cite{hogg99}.

\newpage


\section{One-Step Distance Methods}

In this section we examine the main methods for one-step Hubble constant determination using
astrophysical objects, together with their associated problems and assess the observational
situation with respect to each. Other methods have been proposed 
\epubtkFootnote{For example, one topic that may merit more than a footnote in the future is the study of 
cosmology using gravitational waves. In particular, a coalescing binary system consisting 
of two neutron stars produces gravitational waves, and under those circumstances the 
measurement of the amplitude and frequency of the waves determines the distance to the
object independently of the stellar masses~\cite{schut86}. This was studied in more
detail by~\cite{chern93} and extended to more massive black-hole systems~\cite{holz05,dalal06}. More massive coalescing signals produce lower-frequency 
gravitational wave signals which can be detected with the proposed LISA space-based 
interferometer (\url{http://lisa.nasa.gov/documentation.html}).
The major difficulty is obtaining the redshift measurement to go with
the distance estimate, since the galaxy in which the coalescence event has taken place must 
be identified. Given this, however, the precision of the $H_0$ measurement is limited only 
by weak gravitational lensing along the line of sight, and even this is reducible by 
observations of multiple systems or detailed investigations of matter along the line of
sight. $H_0$ determinations to $\sim$~2\% should be possible, but depend on the launch
of LISA or a similar mission. This is an event which is probably decades away, although
a pathfinder mission to test some of the technology is due for launch in 2015.} but
do not yet have the observations needed to apply them.


\subsection{Megamaser cosmology}
\label{sec:megamaser-cosmology}

To determine the Hubble constant, measurements of distance are
needed. In the nearby universe, the ideal object is one which is
distant enough for peculiar velocities to be small -- in practice
around 50~Mpc -- but for which a distance can be measured in one step
and without a ladder of calibration involving other measurements in
more nearby systems. Megamaser systems in external galaxies offer an
opportunity to do this.

A megamaser system in a galaxy involves clumps of gas which are
typically located $\sim$~0.1~pc from the centre of the
galaxy, close to the central supermassive black hole which is thought
to lie at the centre of most if not all galaxies. These clumps radiate
coherently in the water line at a frequency of approximately
22~GHz. This can be observed at the required milliarcsecond resolution
scale using Very Long Baseline Interferometry (VLBI) techniques. With VLBI
spectroscopy, the velocity  of each individual clump can be measured
accurately, and by repeated observations the movements of each clump
can be followed and the acceleration determined. Assuming that the
clumps are in Keplerian rotation, the radius of each clump from the
central black hole can therefore be calculated, and the distance to
the galaxy follows from knowledge of this radius together with the
angular separation of the clump from the galaxy centre. The black-hole
mass is also obtained as a by-product of the analysis. The analysis is
not completely straightforward, as the disk is warped and viscous,
with four parameters (eccentricity, position angle, periapsis angle
and inclination) describing the global properties of the disk and four
further parameters describing the properties of the
warping~\cite{humph13}. In principle it is vulnerable to systematics
involving the modelling parameters not adequately describing the disk,
but such systematics can be simulated for plausible extra dynamical
components~\cite{humph13} and are likely to be small.

The first maser system to be discovered in an external galaxy was that
in the object NGC~4258. This galaxy has a shell of masers which are
oriented almost edge-on~\cite{miyos95,green95} and apparently in
Keplerian rotation. Measurements of the distance to this galaxy have
become steadily more accurate since the original
work~\cite{herrn99,humph05,humph13}, although the distance of
$\sim 7 \mathrm{\ Mpc}$ to this object is not sufficient to avoid large
(tens of percent) systematics due to peculiar velocities in any attempt to determine
$H_0$.

More recently, a systematic programme has been carried out to
determine maser distances to other, more distant galaxies; the
Megamaser Cosmology Project~\cite{reid09}. The first fruits of this
programme include the measurement of the dynamics of the maser system
in the galaxy UGC~3789, which have become steadily more accurate as
the campaign has progressed~\cite{reid09,braat10,reid13}. A distance
of $49.6 \pm 5.1\mathrm{\ Mpc}$ is determined, corresponding to $H_0 =
68.9 \pm 7.1\kmsmpc$~\cite{reid13};
the error is dominated by the uncertainty in the likely peculiar
velocity, which itself is derived from studies of the Tully--Fisher
relation in nearby clusters~\cite{maste06}. Efforts are under way to
find more megamasers to include in the sample, with success to date in
the cases of NGC~6264 and Mrk~1419. \cite{braat13} and~\cite{kuo13}
report preliminary  results in the cases of the latter two objects,
resulting in an overall determination of $H_0 = 68.0 \pm 4.8\kmsmpc$
($68 \pm 9\kmsmpc$ for NGC~6264). Tightening of the error bars as more
megamasers are discovered, together with careful modelling, are likely
to allow this project to make the cleanest determination of the Hubble
constant within the next five years.

\subsection{Gravitational lenses}
\label{sec:gravitational-lenses}

A general review of gravitational lensing is given by
Wambsganss~\cite{wambs01}; here we review the theory necessary for an
understanding of the use of lenses in determining the Hubble
constant. This determination, like the megamaser method, is a one-step
process, although at a much greater distance. It is thus interesting
both as a complementary determination and as an opportunity to
determine the Hubble parameter as a function of redshift. It has the
drawback of possessing one serious systematic error associated with 
contaminating astrophysics, namely the detailed mass model of the lens.


\subsubsection{Basics of lensing}

Light is bent by the action of a gravitational field. In the case where
a galaxy lies close to the line of sight to a background quasar, the
quasar's light may travel along several different paths to the observer,
resulting in more than one image.

The easiest way to visualise this is to begin with a
zero-mass galaxy (which bends no light rays) acting as the lens, 
and considering all possible 
light paths from the quasar to the observer which have a bend in the
lens plane. From the observer's point of view, we can connect all 
paths which take the same time to reach the observer with a contour
in the lens plane, which in this case is circular in shape. 
The image will form at the centre of the
diagram, surrounded by circles representing increasing light travel
times. This is of course an application of Fermat's principle; images
form at stationary points in the Fermat surface, in this case at the
Fermat minimum. Put less technically, the light has taken a
straight-line path\epubtkFootnote{Strictly speaking, provided we
ignore effects to do with curvature of the Universe.} between the 
source and observer.

If we now allow the galaxy to have a steadily increasing mass, we
introduce an extra time delay (known as the Shapiro delay) along light
paths which pass through the lens plane close to the galaxy centre. This
makes a distortion in the Fermat surface (Figure~\ref{figure_06A}).
 At first, its only effect is
to displace the Fermat minimum away from the distortion. Eventually,
however, the distortion becomes big enough to produce a maximum at the 
position of the galaxy, together with a saddle point on the other side
of the galaxy from the minimum. By Fermat's principle, two further
images will appear at these two stationary points in the Fermat surface.
This is the basic three-image lens configuration, although in practice
the central image at the Fermat maximum is highly demagnified and not
usually seen.

If the lens is significantly elliptical and the lines of sight are well
aligned, we can produce five images, consisting of four images around a
ring alternating between maxima and saddle points, and a central, highly
demagnified Fermat maximum. Both four-image and two-image systems
(``quads'' and ``doubles'') are in fact seen in practice. The major use
of lens systems is for determining mass distributions in the lens
galaxy, since the positions and fluxes of the images carry
information about the gravitational potential of the lens. Gravitational
lensing has the advantage that its effects are independent of whether
the matter is light or dark, so in principle the effects of both
baryonic and non-baryonic matter can be probed.

\epubtkImage{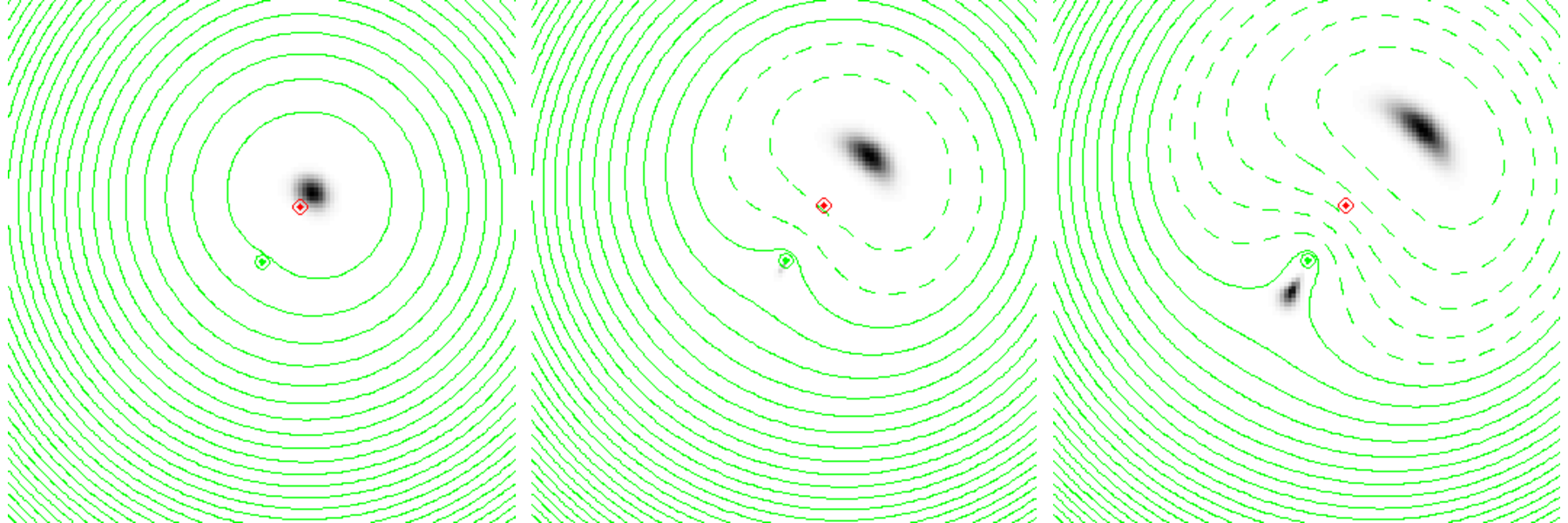}{%
  \begin{figure}[htbp]
    \centerline{\includegraphics[scale=0.2]{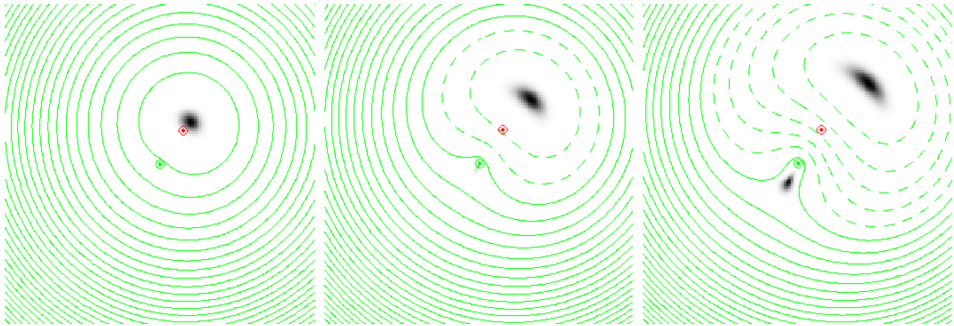}}
    \caption{Illustration of a Fermat surface for a source (red symbol) close
to the line of sight to a galaxy (green symbol). In each case the appearance of
the images to the observer is shown by a greyscale, and the contours of the Fermat surface are
given by green contours. Note that images form at stationary points of the surface
defined by the contours. In the three panels, the mass of the galaxy, and thus the
distortion of the Fermat surface, increases, resulting in an increasingly visible
secondary image at the position of the saddle point. At the same time, the primary
image moves further from the line of sight to the source. In each case the third
image, at the position of the Fermat maximum, is too faint to see.}
    \label{figure_06A}
  \end{figure}}


\subsubsection{Principles of time delays}

Refsdal~\cite{refsd64} pointed out that if the background source is
variable, it is possible to measure an absolute distance within the system and
therefore the Hubble constant. To see how this works, consider the light
paths from the source to the observer corresponding to the individual
lensed images. Although each is at a stationary point in the Fermat time
delay surface, the absolute light travel time for each will generally be
different, with one of the Fermat minima having the smallest travel
time. Therefore, if the source brightens, this brightening will reach
the observer at different times corresponding to the two different light
paths. Measurement of the time delay corresponds to measuring the
difference in the light travel times, each of which is individually
given by
\begin{equation}
  \tau = \frac{D_\mathrm{l} D_\mathrm{s}}{c D_\mathrm{ls}}
  (1 + z_\mathrm{l})
  \left( \frac{1}{2} (\theta - \beta)^2 - \psi (\theta) \right),
  \label{equation_12}
\end{equation}
where $\alpha$, $\beta$ and $\theta$ are angles defined below in
Figure~\ref{figure_06}, $D_\mathrm{l}$, $D_\mathrm{s}$ and 
$D_\mathrm{ls}$ are angular diameter distances also
defined in Figure~\ref{figure_06}, $z_\mathrm{l}$ is the lens redshift, 
and $\psi(\theta)$ is a term representing the Shapiro delay of light 
passing through a gravitational field. Fermat's principle corresponds to 
the requirement that $\nabla\tau=0$. Once the differential time delays 
are known, we can then calculate the ratio of angular diameter distances 
which appears in the above equation. If the source and lens redshifts
are known, $H_0$ follows from Equations~\ref{equation_07B} and~\ref{equation_07C}. 
The value derived depends on the geometric cosmological parameters 
$\Omega_m$ and $\Omega_{\Lambda}$, but this dependence is relatively weak. 
A handy rule of thumb which can be 
derived from this equation for the case of a 2-image lens, if we make 
the assumption that the matter distribution is isothermal\epubtkFootnote{An 
isothermal model is one in which the projected surface mass density 
decreases as $1/r$. An isothermal galaxy will have a flat rotation
curve, as is observed in many galaxies.} and $H_0 = 70\kmsmpc$, is
\begin{equation}
  \Delta \tau = (14 \mathrm{\ days}) (1 + z_\mathrm{l}) D
  \left( \frac{f - 1}{f + 1} \right)s^2,
\end{equation}
where $z_\mathrm{l}$ is the lens redshift, $s$ is the separation of
the images (approximately twice the Einstein radius), $f > 1$ is
the ratio of the fluxes and $D$ is the value of
$D_\mathrm{s}D_\mathrm{l}/D_\mathrm{ls}$ in Gpc. A larger time delay
implies a correspondingly lower $H_0$.

\epubtkImage{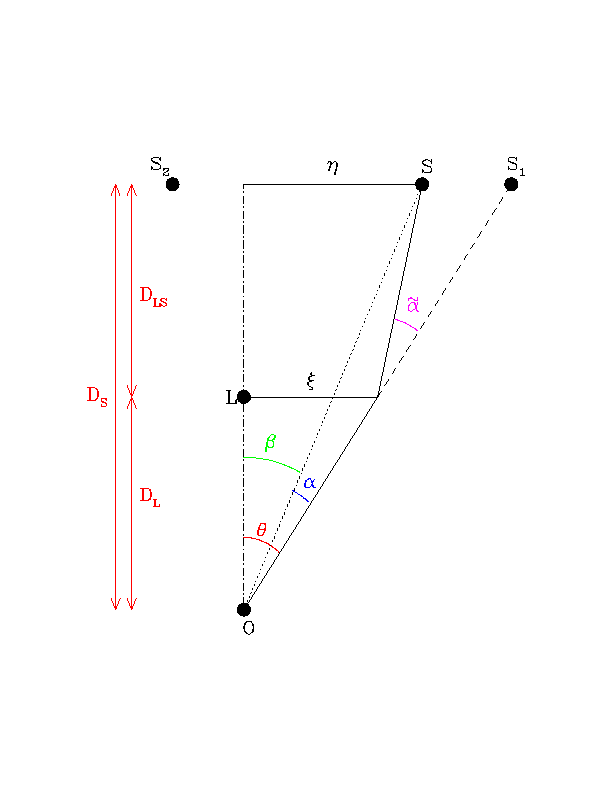}{%
  \begin{figure}[htbp]
    \centerline{\includegraphics[scale=0.6]{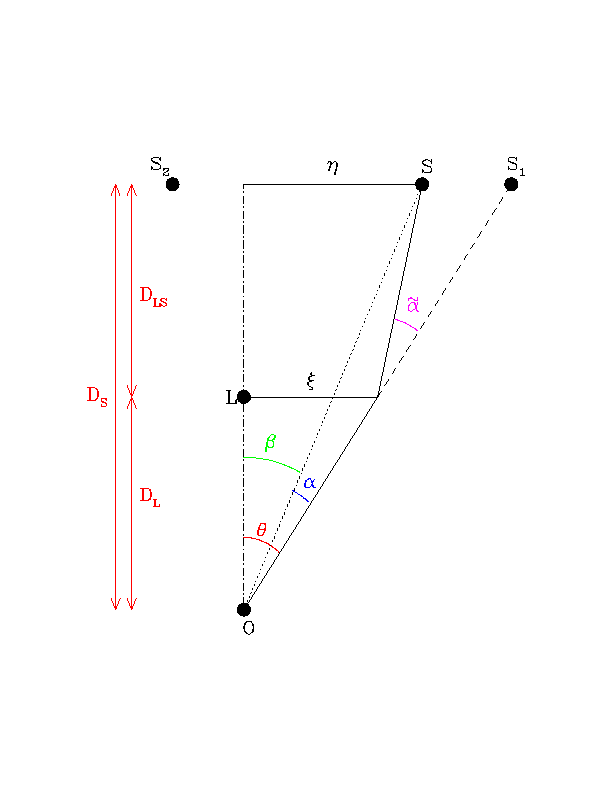}}
    \caption{Basic geometry of a gravitational lens system, reproduced
      from~\cite{wambs01}.}
    \label{figure_06}
  \end{figure}}

The first gravitational lens was discovered in 1979~\cite{walsh79}
and monitoring programmes began soon afterwards to
determine the time delay. This turned out to be a long process involving
a dispute between proponents of a $\sim$~400-day and a
$\sim$~550-day delay, and ended with a determination of
417~\textpm~2~days~\cite{kundi97,schil97}. Since that time,
over 20 more time delays have been determined (see
Table~\ref{table_01}). In the early days, many of the time delays 
were measured at radio wavelengths by examination of those systems in 
which a radio-loud quasar was the multiply imaged source (see 
Figure~\ref{figure_07}). Recently, 
optically-measured delays have dominated, due to the fact that only a small optical
telescope in a site with good seeing is needed for the photometric 
monitoring, whereas radio time delays require large amounts of time on 
long-baseline interferometers which do not exist in large
numbers.\epubtkFootnote{Essentially all radio time delays have come from the
VLA, although 
monitoring programmes with MERLIN have also been
attempted.} A time delay using 
$\gamma$-rays has been determined for one lens~\cite{cheun14} using correlated variations 
in a light-curve which contains emission from both images of the lens.

\epubtkImage{h0_0218.png}{%
  \begin{figure}[htbp]
    \centerline{\includegraphics[width=12cm]{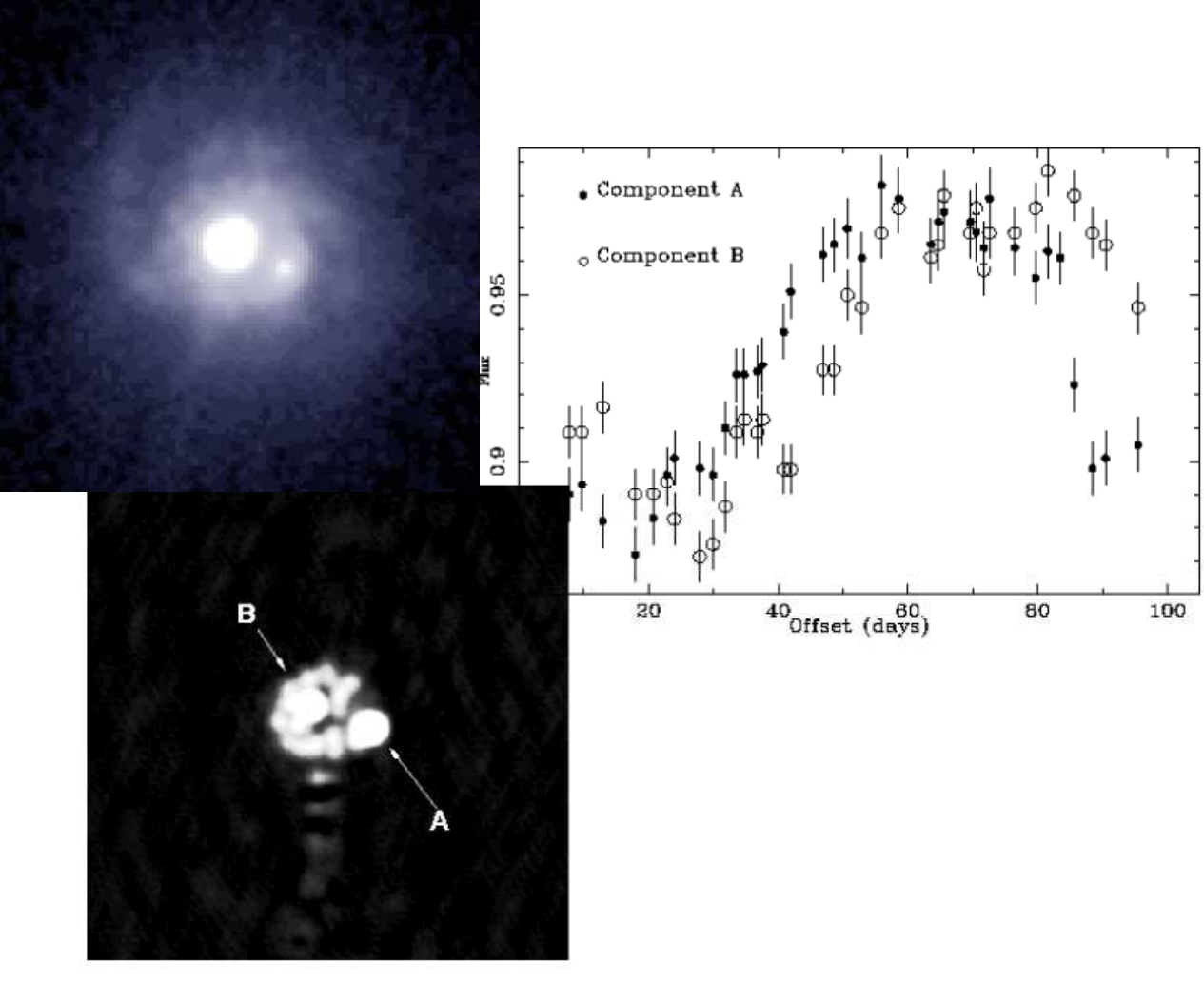}}
    \centerline{\includegraphics[width=14cm]{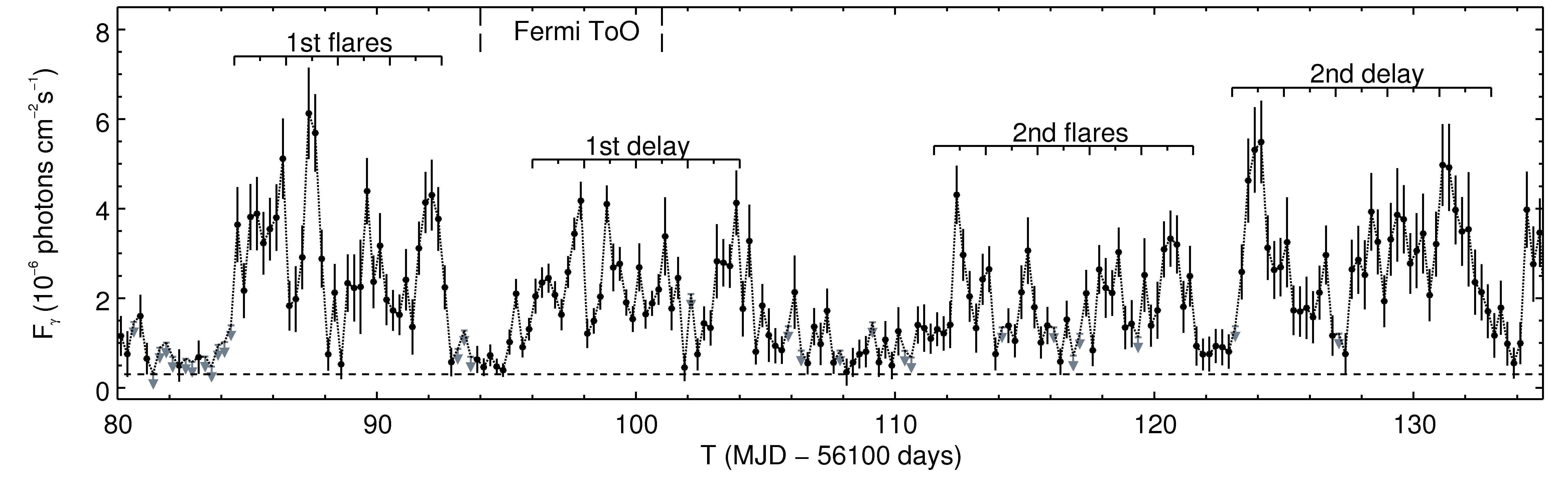}}
    \caption{The lens system JVAS B0218+357. Top right: the measurement
      of time delay of about 10 days from asynchronous variations of
      the two lensed images \cite{biggs99}. The upper left panels show the
      HST/ACS image~\cite{york05} on which can be seen the two images
      and the spiral lensing galaxy, and the radio MERLIN+VLA
      image~\cite{biggs01} showing the two images together with an
      Einstein ring. The bottom panel shows the $\gamma$-ray
      lightcurve~\cite{cheun14}, in which, although the components are
      not resolved, the sharpness of the variations allows a time
      delay to be determined (at 11.46~\textpm~0.16~days, significantly
      greater than the radio time delay).}
    \label{figure_07}
\end{figure}}


\subsubsection{The problem with lens time delays}

Unlike local distance determinations (and even unlike cosmological 
probes which typically use more than one measurement), there is only 
one major systematic piece of astrophysics in the determination of 
$H_0$ by lenses, but it is a very important one.\epubtkFootnote{The redshifts 
of the lens and source also need to be known, as does the position of the centre of the 
lens galaxy; this measurement is not always a trivial proposition\cite{york05}.} This is 
the form of 
the potential in Equation~(\ref{equation_12}). If one parametrises the potential in the form of a power law in projected mass density versus radius, the index
is $-1$ for an isothermal model. This index has a pretty direct
degeneracy\epubtkFootnote{As discussed extensively in~\cite{kocha04b,kocha04a}, 
this is not a global degeneracy, but 
arises because the lensed images tell you about the mass distribution in 
the annulus centred on the galaxy and with inner and outer radii defined
by the inner and outer images. Kochanek~\cite{kocha04b} derives detailed
expressions for the time delay in terms of the central underlying
and controlling parameter, the surface density in this
annulus~\cite{goren88}.}
with the deduced length scale and therefore the Hubble constant; for a
change of 0.1, the length scale changes by about 10\%. The sense of the
effect is that a steeper index, which corresponds to a more centrally
concentrated mass distribution, decreases all the length scales and
therefore implies a higher Hubble constant for a given time delay. 

If an uncertainty in the slope of a power-law mass distribution were the only issue,
then this could be constrained by lensing observables in the case where the source
is extended, resulting in measurements of lensed structure at many different points 
in the lens plane~\cite{kocha01}. This has been done, for example, using multiple radio 
sources~\cite{cohn01}, VLBI radio structure~\cite{wuckn04} and in many objects using 
lensed structure of background galaxies~\cite{bolto06}, although in this latter case
$H_0$ is not measurable because the background objects are not variable. The 
degeneracy between the Hubble constant and the mass model is more general than this, 
however~\cite{goren88}. The reason is that lensing observables give information about 
the derivatives of the Fermat surface; the positions of the images are determined by
the first derivatives of the surface, and the fluxes by the second derivatives. For
any given set of lensing observables, we can move the intrinsic source position,
thus changing the Fermat surface, and then restore the observables to their original
values by adjusting the mass model and thus returning the Fermat surface to its
original configuration. It therefore follows that any given set of measurements of
image positions and fluxes in a lens system is consistent with a number of different
mass models, and therefore a number of different values of $H_0$, because the source 
position cannot be determined. Therefore the assumption of a particular type 
of model, such as a power-law, itself constitutes a selection of a particular one 
out of a range of possible models~\cite{schne13}, each of which would give a 
different $H_0$. Modelling degeneracies arise not only from the mass distribution 
within the lens galaxy, but also from matter along the line of sight. These operate 
in the sense that, if a mass sheet is present which is not known about, the length 
scale obtained is too short and consequently the derived value of $H_0$ is too high.


There are a number of approaches to this mass-degeneracy problem. The first is to
use a non-parametric model for the projected mass distribution, imposing only a
minimum number of physically-motivated requirements such as monotonicity, and
thereby generate large numbers of mass models which are exactly consistent with
the data. This was pioneered by Saha and Williams in a series of papers
\cite{saha97,willi00,saha01,saha06a} in which pixellated models of galaxy mass 
distributions were used. Although pixellated models are useful
for exploring the space of allowed models, they do not break the essential degeneracy.
Other priors may be used, however: in principle it should also be possible to reject 
some possible mass distributions on physical grounds, because we expect the
mass profiles to contain a central stellar cusp and a more extended dark matter halo.  
Undisturbed dark matter haloes should have profiles similar to a Navarro, Frenk \& White 
(NFW,~\cite{navar96}) form, but they may be modified by adiabatic contraction during
the process of baryonic infall when the galaxy forms.

Second, it is possible to increase the reliability of individual lens mass models 
by gathering extra information which partially breaks the mass degeneracy. A major 
improvement is available by the use of stellar velocity dispersions 
\cite{treu02a,treu02b,treu04,koopm06} measured in the lensing galaxy. As a standalone 
determinant of mass models in galaxies at $z\sim0.5$, typical of lens galaxies, such 
measurements are not very useful as they suffer from severe degeneracies with the 
structure of stellar orbits. However, the combination of lensing information (which
gives a very accurate measurement of mass enclosed by the Einstein radius) and 
stellar dynamics (which gives, more or less, the mass enclosed within the effective 
radius of the stellar light) gives a measurement that in effect selects only some
of the family of possible lens models which fit a given set of lensing observables.
The method has large error bars, in part due to residual dependencies on the shape 
of stellar orbits, but also because these measurements are very difficult; each 
galaxy requires about one night of good seeing on a 10-m telescope. Nevertheless, 
this programme has the potential beneficial effect of reducing the dominant systematic 
error, despite the potential additional systematic from the assumptions about stellar
orbits.

Third, we can remove problems associated with mass sheets associated with material
extrinsic to the main lensing galaxy by measuring them using detailed studies of 
the environments of lens galaxies. Studies of lens groups 
\cite{fassn02c,keeto04,fassn06,momch06} show that neglecting matter along the line 
of sight typically has an effect of 10\,--\,20\%, with matter close to the redshift of 
the lens contributing most. More recently, it has been shown that a combination
of studies of number counts and redshifts of nearby objects to the main lens galaxy,
coupled with comparisons to large numerical simulations of matter such as the
Millenium Simulation, can reduce the errors associated with the environment to
around 3\,--\,4\%~\cite{green13}.


\subsubsection{Time delay measurements}

Table~\ref{table_01} shows the currently measured time delays, with references and comments. The
addition of new measurements is now occurring at a much faster rate, due to the
advent of more systematic dedicated monitoring programmes, in particular that
of the COSMOGRAIL collaboration (e.g.~\cite{vuiss07,vuiss08,courb11,rathn13,eulae13}).
Considerable patience is needed for these efforts in
order to determine an unambiguous delay for any given object, given the contaminating
effects of microlensing and also the unavoidable gaps in the monitoring schedule 
(at least for optical monitoring programmes) once per year as the objects move
into the daytime. Derivation of time delays under these circumstances is not a
trivial matter, and algorithms which can cope with these effects have been under
continuous development for decades~\cite{pelt96,kocha04c,hojja13,tewes13} culminating
in a blind analysis challenge~\cite{doble13}.

\begin{table}
  \caption[Time delays, with 1-$\sigma$ errors, from the
    literature.]{Time delays, with 1-$\sigma$ errors, from the
    literature. In some cases multiple delays have been measured in
    4-image lens systems, and in this case each delay is given
    separately for the two components in brackets. An additional time
    delay for CLASS~B1422+231~\cite{patna01}  probably requires
    verification, and a published time delay for Q0142$-$100
    \cite{kopte10,oscoz13} has large errors. Time delays for the CLASS
    and PKS objects have been obtained using radio interferometers,
    and the remainder using optical telescopes.}
  \label{table_01}
  \renewcommand{\arraystretch}{1.3}
  \centering
  \begin{tabular}{l@{\qquad}l@{\qquad}c}
   \toprule
    Lens system & \multicolumn{1}{c}{Time delay} & Reference \\
    & \multicolumn{1}{c}{[days]} \\
   \midrule
    CLASS 0218+357 & \phantom{\qquad}$ 10.5 \pm 0.2 $ & \cite{biggs99} \\
    HE 0435-1-223  & \phantom{\qquad}$ 14.4^{+0.8}_{-0.9} $ (AD) & \cite{kocha06} \\
                   & \phantom{\qquad}$ 7.8 \pm 0.8 $ (BC) & also others \cite{courb11} \\
    SBS 0909+532   & \phantom{\qquad}$ 45^{+1}_{-11} $ ($ 2 \sigma $) & \cite{ullan06} \\
    RX 0911+0551   & \phantom{\qquad}$ 146 \pm 4 $ & \cite{hjort02}\\
    FBQ 0951+2635  & \phantom{\qquad}$ 16 \pm 2 $ & \cite{jakob05} \\
    Q 0957+561     & \phantom{\qquad}$ 417 \pm 3 $ & \cite{kundi97} \\
    SDSS 1001+5027 & \phantom{\qquad}$ 119.3 \pm 3.3 $ & \cite{rathn13} \\
    SDSS 1004+4112 & \phantom{\qquad}$ 38.4 \pm 2.0 $ (AB) & \cite{fohlm06} \\
    SDSS 1029+2623 & & \cite{fohlm13} \\
    HE 1104--185   & \phantom{\qquad}$ 161 \pm 7 $ & \cite{ofek03} \\
    PG 1115+080    & \phantom{\qquad}$ 23.7 \pm 3.4 $ (BC) & \cite{schec97} \\
                   & \phantom{\qquad}$ 9.4 \pm 3.4 $ (AC) & \\
    RX 1131--1231  & \phantom{\qquad}$ 12.0^{+1.5}_{-1.3} $ (AB) & \cite{morga06} \\
                   & \phantom{\qquad}$ 9.6^{+2.0}_{-1.6} $ (AC) & \\
                   & \phantom{\qquad}$ 87 \pm 8 $ (AD) & \\
                   & & \cite{tewes13} \\
    SDSS J1206+4332 & \phantom{\qquad}$ 111.3 \pm 3 $ & \cite{eulae13} \\
    SBS 1520+530   & \phantom{\qquad}$ 130 \pm 3 $ & \cite{burud02a} \\
    CLASS 1600+434 & \phantom{\qquad}$ 51 \pm 2 $ & \cite{burud00} \\
                   & \phantom{\qquad}$ 47^{+5}_{-6} $ & \cite{koopm00} \\
    CLASS 1608+656 & \phantom{\qquad}$ 31.5^{+2}_{-1} $ (AB) & \cite{fassn02a} \\
                   & \phantom{\qquad}$ 36^{+1}_{-2} $ (BC) & \\
                   & \phantom{\qquad}$ 77^{+2}_{-1} $ (BD) & \\
    SDSS 1650+4251 & \phantom{\qquad}$ 49.5 \pm 1.9 $ & \cite{vuiss07} \\
    PKS 1830--211  & \phantom{\qquad}$ 26^{+4}_{-5} $ & \cite{lovel98} \\
    WFI J2033--4723 & \phantom{\qquad}$ 35.5 \pm 1.4 $ (AB) & \cite{vuiss08} \\
    HE 2149--2745  & \phantom{\qquad}$ 103 \pm 12 $ & \cite{burud02b} \\
    HS 2209+1914   & \phantom{\qquad}$ 20.0 \pm 5 $ & \cite{eulae13} \\
    Q 2237+0305    & \phantom{\qquad}$ 2.7^{+0.5}_{-0.9} \mathrm{\ h} $ & \cite{dai03} \\
\bottomrule
\end{tabular}
\end{table}

\subsubsection{Derivation of \textit{H}\sub{0}: Now, and the future}

Initially, time delays were usually turned into Hubble constant values using assumptions 
about the mass model -- usually that of a single, isothermal power law~\cite{koopm06} -- 
and with rudimentary modelling of the environment of the lens system as necessary. Early 
analyses of this type resulted in rather low values of the Hubble constant~\cite{kocha02}
for some systems, 
sometimes due to the steepness of the lens potential~\cite{treu02a}. As the number of 
measured time delays expanded, combined analyses of multiple lens systems were conducted, 
often assuming parametric lens models~\cite{oguri07} but also using Monte Carlo methods to 
account for quantities such as the presence of clusters around the main lens. These methods 
typically give values around $70\kmsmpc$ (e.g., ($68 \pm 6 \pm 8)\kmsmpc$ from Oguri 2007), but 
with an uncomfortably greater spread between lens systems than would be expected on the 
basis of the formal errors. An alternative approach to composite modelling is to use 
non-parametric lens models, on the grounds that these may permit a wider range of mass 
distributions~\cite{saha06a,paraf10} even though they also contain some level of prior
assumptions. Saha et al.~(2006) used ten time-delay lenses for 
this purpose, and Paraficz et al.~(2010) extended the analysis to eighteen systems
obtaining $66^{+6}_{-4}\kmsmpc$ , with  a further extension by Sereno \& Paraficz in 2014
\cite{seren14} giving $66 \pm 6 \pm 4$ (stat/syst) \kmsmpc.

In the last few years, concerted attempts have emerged to put together improved
time-delay observations with systematic modelling. For two existing time-delay
lenses (CLASS~B1608+656 and RXJ~1131--1231) modelling has been undertaken 
\cite{suyu10,suyu13a} using a combination of all of the previously described 
ingredients: stellar velocity dispersions to constrain the lens model and partly
break the mass degeneracy, multi-band HST imaging to evaluate and model the 
extended light distribution of the lensed object, comparison with numerical 
simulations to gauge the likely contribution of the line of sight to the lensing
potential, and the performance of the analysis blind (without sight of the 
consequences for $H_0$ of any decision taken during the modelling). The results
of the two lenses together, $75.2^{+4.4}_{-4.2}$ and $73.1^{+2.4}_{-3.6}\kmsmpc$ in
flat and open $\Lambda$CDM, respectively, are probably the most reliable 
determinations of $H_0$ from lensing to date, even if they do not have the lowest
formal error\epubtkFootnote{The programme, known as H0LiCOW, is now continuing in
order to measure time delays, improve models and derive further $H_0$ values for 
more lenses.}.

In the immediate future, the most likely advances come from further analysis of
existing time delay lenses, although the process of obtaining the data for good
quality time delays and constraints on the mass model is not a quick process. A
number of further developments will expedite the process. The first is the likely
discovery of lenses on an industrial scale using the Large Synoptic Survey 
Telescope (LSST,\cite{ivezi08}) and the Euclid satellite~\cite{amend13}, 
together with time delays produced by 
high cadence monitoring. The second is the availability in a few years' time of
$>$~8-m class optical telescopes, which will ease the followup problem considerably.
A third possibility which has been discussed in the past is the use of
double source-plane lenses, in which two background objects, one of
which is a quasar, are imaged by a single foreground object \cite{gavaz08,colle12}.
Unfortunately it appears \cite{schne14} that even this additional set of constraints
leave the mass degeneracy intact, although it remains to be seen whether dynamical
information will help relatively more in these objects than in single-plane systems.

One potentially clean way to break mass model degeneracies is to
discover a lensed type Ia supernova~\cite{oguri03a,oguri03b}. 
The reason is that, as we have seen, the intrinsic 
brightness of SNe~Ia can be determined from their lightcurve, and it can be 
shown that the resulting absolute magnification of the images can then 
be used to bypass the effective degeneracy between the Hubble constant 
and the radial mass slope. Oguri et~al.~\cite{oguri03b} and also 
Bolton and Burles~\cite{bolto03} discuss prospects for finding such objects; 
future surveys with the Large Synoptic Survey Telescope (LSST) are likely 
to uncover significant numbers of such 
events. The problem is likely to be the determination of the time
delay, since nearly all such objects are subject to significant
microlensing effects within the lensing galaxy which is likely to
restrict the accuracy of the measurement~\cite{doble06}.


\subsection{The Sunyaev--Zel'dovich effect}

The basic principle of the Sunyaev--Zel'dovich (S-Z) method~\cite{sunya72},
including its use to determine the Hubble constant~\cite{silk78}, is
reviewed in detail in~\cite{birki99,carls02}. It is based on the physics
of hot ($10^8 \mathrm{\ K}$) gas in clusters, which emits X-rays by
bremsstrahlung emission with a surface brightness given by the equation
\begin{equation}
  b_{\mathrm{X}} = \frac{1}{4 \pi (1 + z)^3} \int n_e^2 \Lambda_e \, dl
\end{equation}
(see e.g.~\cite{birki99}), where $n_e$ is the electron density and 
$\Lambda_e$ the spectral emissivity, which depends on the electron
temperature. 

At the same time, the electrons of the hot gas in the cluster Compton
upscatter photons from the CMB radiation. At radio frequencies below the
peak of the Planck distribution, this causes a ``hole'' in radio
emission as photons are removed from this spectral region and turned
into higher-frequency photons (see Figure~\ref{figure_08}). The
decrement is given by an optical-depth equation,
\begin{equation}
\Delta I (\nu) = I_0 \int n_e \sigma_T \Psi(\nu, T_e) \, dl,
\end{equation}
involving many of the same parameters and a function $\Psi$
which depends on frequency and electron temperature. 
It follows that, if both $b_{\mathrm{X}}$
and $\Delta I(x)$ can be measured, we have two equations for the
variables $n_e$ and the integrated length $l_{\parallel}$ through the
cluster and can calculate both quantities. Finally, if we assume that
the projected size $l_{\perp}$ of the cluster on the sky is equal to
$l_{\parallel}$, we can then derive an angular diameter distance if we
know the angular size of the cluster. The Hubble constant is then easy
to calculate, given the redshift of the cluster.

\epubtkImage{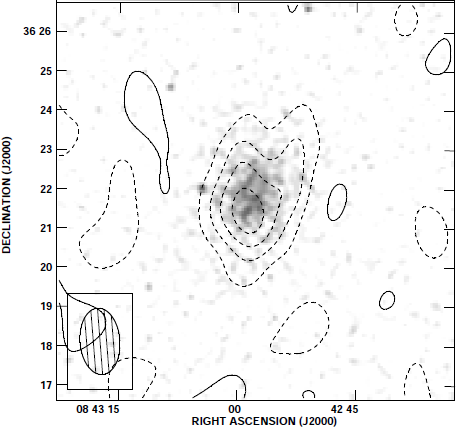}{%
  \begin{figure}[htbp]
    \centerline{\includegraphics[scale=0.7]{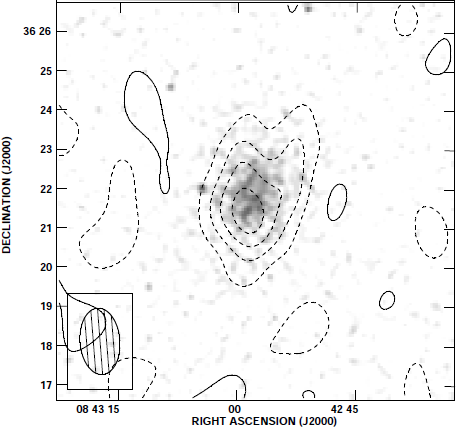}}
    \caption{S-Z decrement observation of Abell 697 with the Ryle
      telescope in contours superimposed on the ROSAT grey-scale
      image. Reproduced from~\cite{jones05}.}
    \label{figure_08}
  \end{figure}}

\begin{table}
  \caption[Some recent measurements of $H_0$ using the
    S-Z effect.]{Some recent measurements of $H_0$ using the
    S-Z effect. Model types are $\beta$ for the assumption of a
    $\beta$-model and H for a hydrostatic equilibrium model. Some of the studies target
the same clusters, with three objects being common to more than one of the four smaller
studies, The larger study \cite{bonam06} contains four of the objects from \cite{jones05} 
   and two from \cite{schmi04}.}
  \label{table_02}
  \renewcommand{\arraystretch}{1.3}
  \centering
  \begin{tabular}{l@{\qquad}c@{\qquad}c@{\qquad}l}
    \toprule
    Reference & Number of clusters & Model type &
    \multicolumn{1}{c}{$H_0$ determination} \\
    & & &
    \multicolumn{1}{c}{[$ \mathrm{km\ s}^{-1}\mathrm{\ Mpc}^{-1}$]} \\
    \midrule
    \cite{bonam06} & $ 38 $ & $ \beta + \mathrm{H} $ &
    \phantom{\qquad}$ 76.9^{+3.9 +10.0}_{-3.4 -8.0} $ \\
    \cite{jones05} & $ 5 $  & $ \beta $              &
    \phantom{\qquad}$ 66^{+11 +9}_{-10 -8} $ \\
    \cite{udomp04} & $ 7 $  & $ \beta $              &
    \phantom{\qquad}$ 67^{+30 +15}_{-18 -6} $ \\
    \cite{schmi04} & $ 3 $  & $ \mathrm{H} $         &
    \phantom{\qquad}$ 69 \pm 8 $ \\
    \cite{mason01} & $ 7 $  & $ \beta $              &
    \phantom{\qquad}$ 66^{+14 +15}_{-11 -15} $ \\
    \cite{reese02} & $ 18 $ & $ \beta $              &
    \phantom{\qquad}$ 60^{+4 +13}_{-4 -18} $ \\
    \bottomrule
  \end{tabular}
\end{table}

Although in principle a clean, single-step method, in practice there are
a number of possible difficulties. Firstly, the method involves two
measurements, each with a list of possible errors. The X-ray
determination carries a calibration uncertainty and an uncertainty due
to absorption by neutral hydrogen along the line of sight. The radio
observation, as well as the calibration, is subject to possible errors
due to subtraction of radio sources within the cluster which are
unrelated to the S-Z effect. Next, and probably most importantly, are the
errors associated with the cluster modelling. In order to extract
parameters such as electron temperature, we need to model the physics of
the X-ray cluster. This is not as difficult as it sounds, because X-ray
spectral information is usually available, and line ratio measurements
give diagnostics of physical parameters. For this modelling the cluster is
usually assumed to be in hydrostatic equilibrium, or a ``beta-model'' (a
dependence of electron density with radius of the form 
$n(r)=n_0(1+r^2/r_\mathrm{c}^2)^{-3\beta/2}$) is assumed. Several
recent works~\cite{schmi04,bonam06} relax this assumption, 
instead constraining the profile of the cluster with available X-ray 
information, and the dependence of $H_0$ on these details is
often reassuringly small ($< 10\%$). Finally, the cluster selection can be done
carefully to avoid looking at prolate clusters along the long axis 
(for which $l_{\perp}\neq l_{\parallel}$)
and therefore seeing more X-rays than one would predict. This can be
done by avoiding clusters close to the flux limit of X-ray flux-limited
samples, Reese et~al.~\cite{reese02} estimate an overall
random error budget of 20\,--\,30\% for individual clusters. As in the case
of gravitational lenses, the problem then becomes the relatively trivial
one of making more measurements, provided there are no unforeseen
systematics.

The cluster samples of the most recent S-Z determinations (see
Table~\ref{table_02}) are
not independent in that different authors often observe the same clusters.
The most recent work, that in~\cite{bonam06} is larger than the
others and gives a higher $H_0$. It is worth noting, however, that if
we draw subsamples from this work and compare the results with the other
S-Z work, the $H_0$ values from the subsamples are consistent. For
example, the $H_0$ derived from the data in~\cite{bonam06} and modelling
of the five clusters also considered in~\cite{jones05} is actually 
lower than the value of $66\kmsmpc$ in~\cite{jones05}. Within the smaller
samples, the scatter is much lower than the quoted errors, partially due
to the overlap in samples (three objects are common to more than one of
the four smaller studies).

It therefore seems as though S-Z determinations of the Hubble constant
are beginning to converge to a value of around $70\kmsmpc$, although the
errors are still large, values in the low to mid-sixties are still
consistent with the data and it is possible that some objects may have
been observed but not used to derive a published $H_0$ value. Even more 
than in the case of gravitational
lenses, measurements of $H_0$ from individual clusters are occasionally 
discrepant by factors of nearly two in either direction, and it would
probably teach us interesting astrophysics to investigate these cases
further.

\subsection{Gamma-ray propagation}

High-energy $\gamma$-rays emitted by distant AGN are subject to interactions with ambient 
photons during their passage towards us, producing electron-positron pairs. The mean free 
path for this process varies with photon energy, being smaller at higher energies, and is
generally a substantial fraction of the distance to the sources. The observed spectrum of
$\gamma$-ray sources therefore shows a high-energy cutoff, whose characteristic energy
decreases with increasing redshift. The expected cutoff, and its dependence on redshift,
has been detected with the \textit{Fermi} satellite~\cite{acker12}.

The details of this effect depend on the Hubble constant, and can therefore be used to
measure it~\cite{salam94,barra08}. Because it is an optical depth effect, knowledge of
the interaction cross-section from basic physics, together with the number density $n_p$ 
of the interacting photons, allows a length measurement and, assuming knowledge of the 
redshift of the source, $H_0$. In practice, the cosmological world model is also needed 
to determine $n_p$ from observables. From the existing \textit{Fermi} data a value of
$72\kmsmpc$ is estimated~\cite{domin13} although the errors, dominated by the calculation 
of the evolution of the extragalactic background light using galaxy luminosity functions 
and spectral energy distributions, are currently quite large 
($\sim 10\kmsmpc$).


\newpage


\section{Local Distance Ladder}
\label{sec:local-distance-ladder}

\subsection{Preliminary remarks}

As we have seen, in principle a single object whose spectrum reveals its
recession velocity, and whose distance or luminosity is accurately known,
gives a measurement of the Hubble constant. In practice, the object must
be far enough away for the dominant contribution to the motion to be the
velocity associated with the general expansion of the Universe (the ``Hubble
flow''), as this expansion velocity increases linearly with distance
whereas other nuisance velocities, arising from gravitational
interaction with nearby matter, do not. For nearby galaxies,
motions associated with the potential of the local environment are about
$200\mbox{\,--\,}300 \mathrm{\ km\ s}^{-1}$, requiring us to measure
distances corresponding to recession velocities of a few thousand
$\mathrm{\ km\ s}^{-1}$ or greater. These recession velocities
correspond to distances of at least a few tens of Mpc.

The Cepheid distance method, used since the 
original papers by Hubble, has therefore been to measure distances of 
nearby objects and use this knowledge to calibrate the brightness of more
distant objects compared to the nearby ones. This process must be repeated 
several times in order to bootstrap one's way out to tens of Mpc, and
has been the subject of many reviews and books (see e.g., \cite{rowan85}). 
The process has a long and tortuous history, with many
controversies and false turnings, and which as a
by-product included the discovery of a large amount of stellar
astrophysics. The astrophysical content of the method is a disadvantage,
because errors in our understanding propagate directly into errors in
the distance scale and consequently the Hubble constant. The number of
steps involved is also a disadvantage, as it allows opportunities for
both random and systematic errors to creep into the measurement. It is
probably fair to say that some of these errors are still not universally
agreed on. The range of recent estimates is in the low seventies of
$\mathrm{\ km\ s}^{-1}\mathrm{\ Mpc}^{-1}$, with the errors 
having shrunk by a factor of two in the last ten years, and
the reasons for the disagreements (in many cases by different analysis
of essentially the same data) are often quite complex.


\subsection{Basic principle}

We first outline the method briefly, before discussing each stage in
more detail. Nearby stars have a reliable
distance measurement in the form of the parallax effect. This effect
arises because the earth's motion around the sun produces an apparent 
shift in the position of nearby stars compared to background stars at 
much greater distances. The shift has a period of a year, and an angular
amplitude on the sky of the Earth-Sun distance divided by the distance to
the star. The definition of the parsec is the distance which gives a
parallax of one arcsecond, and is equivalent to 3.26 light-years, or
$3.09 \times 10^{16} \mathrm{\ m}$. The field of parallax measurement was
revolutionised by the Hipparcos satellite, which measured thousands of
stellar parallax distances, including observations of 223 Galactic Cepheids;
of the Cepheids, 26 yielded determinations of reasonable
significance~\cite{feast97}. The GAIA satellite will increase these by
a large factor, probably observing thousands of Galactic Cepheids and
giving accurate distances as well as colours and metallicities~\cite{turon12}.

Some relatively nearby stars exist in clusters of a few hundred stars 
known as ``open clusters''. These stars can be plotted on a
Hertzsprung--Russell diagram of temperature, deduced from their colour
together with Wien's law, against apparent luminosity. Such plots
reveal a characteristic sequence, known as the ``main sequence'' which
ranges from red, faint stars to blue, bright stars. This sequence
corresponds to the main phase of stellar evolution which stars occupy
for most of their lives when they are stably burning hydrogen. In some
nearby clusters, notably the Hyades, we have stars all at the same
distance and for which parallax effects can give the 
absolute distance to $<$1\%~\cite{perry98}. In such cases, 
the main sequence can be calibrated so that we can
predict the absolute luminosity of a main-sequence star of a given
colour. Applying this to other clusters, a process known as ``main
sequence fitting'', can also give the absolute distance to these other
clusters; the errors involved in this fitting process appear to be
of the order of a few percent~\cite{an07}.

The next stage of the bootstrap process is to determine the distance to
the nearest objects outside our own Galaxy, the Large and Small
Magellanic Clouds. For this we can apply the open-cluster method
directly, by observing open clusters in the LMC. Alternatively, we can
use calibrators whose true luminosity we know, or can predict from their
other properties. Such calibrators must be present in the LMC and also
in open clusters (or must be close enough for their parallaxes to be
directly measurable).

These calibrators include Mira variables, RR Lyrae stars and Cepheid
variable stars, of which Cepheids are intrinsically the most luminous. 
All of these have variability periods which are correlated with their 
absolute luminosity (section 1.1), and in principle the
measurement of the distance of a nearby object of any of these types can
then be used to determine distances to more distant similar objects
simply by observing and comparing the variability periods.

The LMC lies at about 50~kpc, about three orders of magnitude 
less than that of the distant galaxies of interest for the Hubble constant. 
However, one class of variable stars, Cepheid variables, can be seen in both 
the LMC and in galaxies at distances up to 20-30~Mpc. The coming of the
Hubble Space Telescope has been vital for this process, as only with the
HST can Cepheids be reliably identified and measured in such 
galaxies. 

\epubtkImage{}{%
  \begin{figure}[htbp]
    \centerline{\includegraphics[width=7cm]{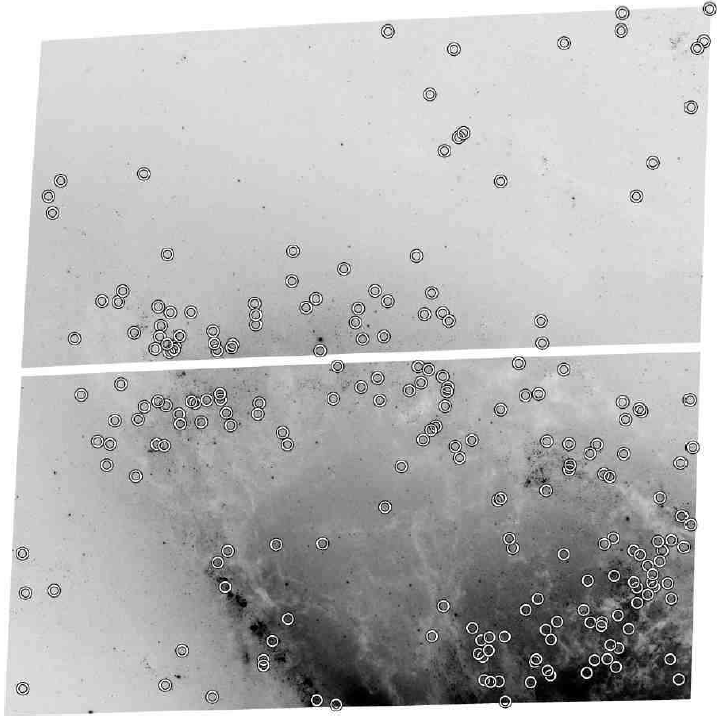}}\vspace{1cm}
    \centerline{\includegraphics[width=12cm]{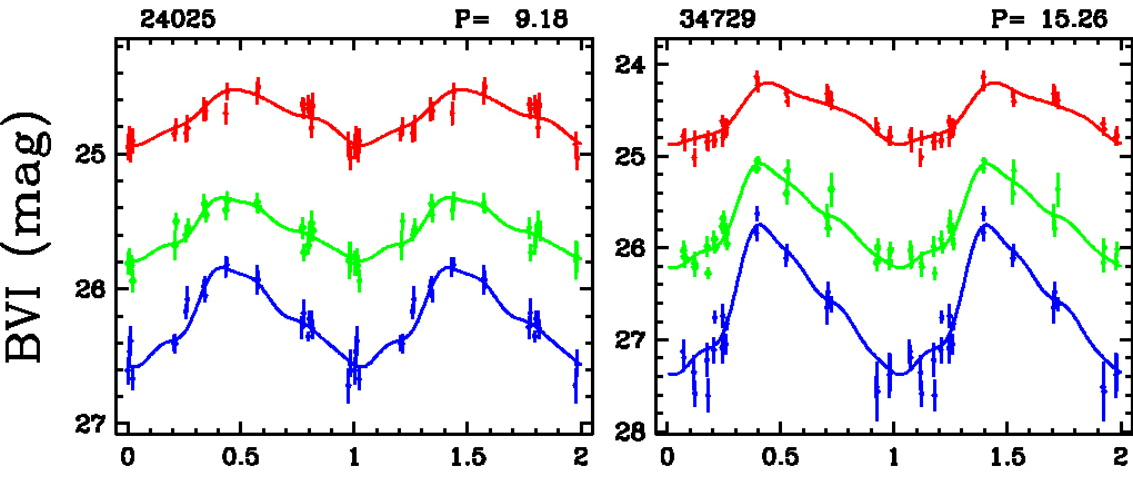}}
    \caption{Positions of Cepheid variables in HST/ACS observations of
      the galaxy NGC~4258, reproduced from~\cite{macri06}. Typical
      Cepheid lightcurves are shown on the top.}
\end{figure}}

Even the HST galaxies containing Cepheids are not sufficient to allow
the measurement of the universal expansion, because they are not distant
enough for the dominant velocity to be the Hubble flow. 
The final stage is to use galaxies with distances measured with
Cepheid variables to calibrate other indicators which can be measured to
cosmologically interesting distances. The most promising indicator
consists of type~Ia supernovae (SNe), which are produced by binary systems
in which a giant star is dumping mass on to a white dwarf which has
already gone through its evolutionary process and collapsed to an
electron-degenerate remnant; at a critical point, the rate and amount of
mass dumping is sufficient to trigger a supernova explosion. The physics
of the explosion, and hence the observed light-curve of the rise and
slow fall, has the same characteristic regardless of distance. Although
the absolute luminosity of the explosion is not constant, type~Ia
supernovae have similar light-curves~\cite{pskov67,barbo73,tamma82}
and in particular there is a
very good correlation between the peak brightness and the degree of
fading of the supernova 15~days\epubtkFootnote{Because of the expansion of the
Universe, there is a time dilation of a factor $(1+z)^{-1}$ which must
be applied to timescales measured at cosmological distances before these
are used for such comparisons.} after peak brightness (a quantity known
as $\Delta m_{15}$~\cite{phill93,hamuy96}). 
If SNe~Ia can be detected in galaxies with known
Cepheid distances, this correlation can be calibrated and used to
determine distances to any other galaxy in which a SN~Ia is detected.
Because of the brightness of supernovae, they can be observed at large
distances and hence, finally, a comparison between redshift and distance
will give a value of the Hubble constant.

There are alternative indicators which can be used instead of SNe~Ia for
determination of $H_0$; all of them rely on the correlation of some
easily observable property of galaxies with their luminosity or size,
thus allowing them to be used as standard candles or rulers respectively. For
example, the edge-on rotation velocity $v$ of spiral galaxies scales with
luminosity as $L\propto v^4$, a scaling known as the Tully--Fisher
relation~\cite{tully77}. There is an equivalent for elliptical 
galaxies, known as the Faber--Jackson relation~\cite{faber76}. 
In practice, more complex combinations of observed properties are often
used such as the $D_n$ parameter of~\cite{dress87} and~\cite{lynde88},
to generate measurable properties of elliptical
galaxies which correlate well with luminosity, or the ``fundamental
plane''~\cite{dress87,djorg87} between three properties,
the average surface brightness within an effective radius\epubtkFootnote{The 
effective radius is the radius from within which half the galaxy's light is emitted.} 
the effective radius $r_e$, and the central stellar velocity
dispersion $\sigma$. Here we can measure surface brightnesses and $\sigma$
and derive a standard ruler in terms of the true $r_e$ which can then be
compared with the apparent size of the galaxy. 

A somewhat different indicator relies on the fact that 
the degree to which stars within galaxies are resolved depends on
distance, in the sense that closer galaxies have more statistical
``bumpiness'' in the surface-brightness distribution~\cite{tonry88}
because of the degree to which Poisson fluctuations in the stellar
surface density are visible. This method of surface brightness fluctuation 
can also be calibrated by Cepheid variables in the nearer galaxies.


\subsection{Problems and comments}


\subsubsection{Distance to the LMC}

The LMC distance is probably the best-known, and least controversial,
part of the distance ladder. Some methods of determination are
summarised in~\cite{feast99}; independent calibrations using RR 
Lyrae variables, Cepheids and open clusters, are consistent with a 
distance of $\sim 50 \mathrm{\ kpc}$. An early measurement, independent of all of
the above, was made by~\cite{panag91} using the
type~II supernova 1987A in the LMC. This supernova produced an expanding
ring whose angular diameter could be measured using the HST. An absolute
size for the ring could also be deduced by monitoring ultraviolet
emission lines in the ring and using light travel time arguments, and
the distance of $51.2 \pm 3.1 \mathrm{\ kpc}$ followed from comparison of the two. An
extension to this light-echo method was proposed in~\cite{spark94} which
exploits the fact that the maximum in polarization in scattered light is
obtained when the scattering angle is $90^\circ$. Hence, if a
supernova light echo were observed in polarized light, its distance
would be unambiguously calculated by comparing the light-echo time and
the angular radius of the polarized ring.

More traditional calibration methods traditionally resulted in distance moduli
to the LMC of $\mu^0 \simeq 18.50$ (defined as $5\log d-5$, where $d$ is the 
distance in parsecs) corresponding to a distance of $\simeq 50 \mathrm{\ kpc}$. In
particular, developments in the use of standard-candle stars, main sequence
fitting and the details of SN~1987A are reviewed by~\cite{alves04} who
concludes that $\mu^0 = 18.50 \pm 0.02$. This has
recently been revised downwards slightly using a more direct calibration
using parallax measurements of Galactic Cepheids~\cite{bened07} to calibrate
the zero-point of the Cepheid \textit{P-L} relation in the LMC~\cite{freed10}. 
A value of $\mu^0=18.40 \pm 0.01$ is found by these authors, corresponding
to a distance of $47.9 \pm 0.2 \mathrm{\ kpc}$. The likely corresponding error in $H_0$ 
is well below the level of systematic errors in other parts of the distance 
ladder. This LMC distance also agrees well with the value needed in order
to make the Cepheid distance to NGC~4258 agree with the maser distance
(\cite{macri06}, see also Section~\ref{sec:CMB}).

\subsubsection{Cepheid systematics}

The details of the calibration of the Cepheid period-luminosity relation
have historically caused the most difficulties in the local calibration of
the Hubble constant. There are a number of minor effects, which can be estimated
and calibrated relatively easy, and a dependence on metallicity which is a
systematic problem upon which a lot of effort has been spent and which is
now considerably better understood.

One example of a minor difficulty is a selection bias in Cepheid programmes; faint
Cepheids are harder to see. Combined with the correlation between
luminosity and period, this means that only the brighter short-period
Cepheids are seen, and therefore that the \textit{P-L} relation in distant
galaxies is made artificially shallow~\cite{sanda88} resulting in
underestimates of distances. Neglect of this bias can give
differences of several percent in the answer, and detailed simulations
of it have been carried out by Teerikorpi and collaborators (e.g.
\cite{teeri02,patur04,patur05,patur06}. Most authors
correct explicitly for this problem -- for example,~\cite{freed01}
calculate the correction analytically and find a maximum bias of
about 3\%. Teerikorpi \& Paturel suggest that a residual bias may still
be present, essentially because the amplitude of variation introduces an
additional scatter in brightness at a given period, in addition to the
scatter in intrinsic luminosity. How big this bias is is hard to
quantify, although it can in principle be eliminated by using only
long-period Cepheids at the cost of increases in the random error.

The major systematic difficulty became apparent in studies of the biggest 
sample of Cepheid variables, which arises 
from the OGLE microlensing survey of the LMC~\cite{udals99}. Samples of
Galactic Cepheids have been studied by many authors
\cite{feast99,giere98,fouqu03,barne03,bened02,kerve04}, and their distances
can be calibrated by the methods previously described, or by using 
lunar-occultation measurements of stellar angular diameters~\cite{fouqu97}
together with stellar temperatures to determine distances by Stefan's
law~\cite{wesse69,barne76}. Comparison of the \textit{P-L} relations for 
Galactic and LMC Cepheids, however, show significant dissimilarities.
In all three HST colours ($B$,$V$,$I$) the slope of the relations are
different, in the sense that Galactic Cepheids are brighter than LMC
Cepheids at long periods and are fainter at short periods. The two 
samples are of equal brightness in $B$ at a period of approximately 30
days, and at a period of a little more than 10 days in $I$.\epubtkFootnote{Nearly
all Cepheids measured in galaxies containing SN~Ia have periods $>20$
days, so the usual sense of the effect is that Galactic Cepheids of a
given period are brighter than LMC Cepheids.}

The culprit for this discrepancy is mainly metallicity\epubtkFootnote{Here, 
as elsewhere in astronomy, the term ``metals'' is used to refer to any 
element heavier than helium. Metallicity is usually quoted as 12+log(O/H), 
where O and H are the abundances of oxygen and hydrogen.} differences in
the Cepheids, which in turn results from the fact that the LMC is more
metal-poor than the Galaxy. Unfortunately, many of the external galaxies
which are to be used for distance determination are likely to be similar
in metallicity to the Galaxy, but the best local information on Cepheids
for calibration purposes comes from the LMC. On average, the Galactic
Cepheids tabulated by~\cite{groen04} are of approximately of solar 
metallicity, whereas those of the LMC are approximately 0.6~dex less
metallic. If these two samples are compared with their independently
derived distances, a correlation of brightness with metallicity appears
with a slope of $-0.8 \pm 0.3 \mathrm{\ mag\ dex}^{-1}$ using only Galactic Cepheids,
and $-0.27 \pm 0.08 \mathrm{\ mag\ dex}^{-1}$ using both samples together. This
can cause differences of 10\,--\,15\% in inferred distance if the effect is
ignored.

Many areas of historic disagreement can be traced back to how this 
correction is done. In particular, two different 2005\,--\,2006 estimates of 
73~\textpm~4\,(statistical)~\textpm~5\,(systematic)\kmsmpc~\cite{riess05} and
62.3~\textpm~1.3\,(statistical)~\textpm~5\,(systematic)\kmsmpc~\cite{sanda06},
both based on the same Cepheid photometry from the HST Key 
Project\cite{saha06} and essentially the same Cepheid \textit{P-L} relation
for the LMC~\cite{thim03} have their origin mainly in this 
effect.\epubtkFootnote{The details are discussed in more detail in an earlier 
version of this review~\cite{jacks07}.} One can apply a global correction
for metallicity differences between the LMC and the galaxies in which
the Cepheids are measured by the HST Key Project~\cite{sakai04}, or
attempt to interpolate between LMC and Galactic \textit{P-L} relations 
\cite{tamma03} using a period-dependent metallicity correction
\cite{sanda06}. The differences in this correction account for the
10\,--\,15\% difference in the resulting value of $H_0$.

More recently, a number of different solutions for this problem have been
found, which are summarised in the review by~\cite{freed10} and many of which
involve getting rid of the intermediate LMC step using other calibrations.
\cite{macri06} use ACS observations of Cepheids in the galaxy NGC~4258,
which has a well-determined distance using maser 
observations (Section 4, \cite{humph08,green09,humph13}), and whose
Cepheids have a range of metallicities\cite{zarit94}. Analysis of these
Cepheids suggests that the use of a \textit{P-L} relation whose slope varies
with metallicity~\cite{tamma03,sanda06} overcorrects at long period. Because
of the known maser distance, these Cepheids can then be used both to 
determine the LMC distance independently~\cite{macri06} and also to 
calibrate the SNe distance scale and hence determine 
$H_0$~\cite{riess09a,riess09b}. The estimate has been incrementally
improved by several methods in the last few years

Values obtained for the Hubble constant using the NGC~4258 calibration 
are quoted by~\cite{riess11}
as $74.8 \pm 3.1\kmsmpc$, using a value of 7.28~Mpc as the NGC~4258 distance.
This was later corrected by~\cite{humph13}, who find a distance of
7.60~\textpm~0.17\,(stat)~\textpm~0.15\,(sys)~Mpc using more VLBI epochs, together
with better modelling of the masers, which therefore yields a
Hubble constant of $72.0 \pm 3.0\kmsmpc$. Efstathiou~\cite{efsta13} has 
argued for further modifications, with different criteria for rejecting
outlying Cepheids; this lowers $H_0$ to $70.6 \pm 3.3\kmsmpc$.
The alternative distance ladder measurement, using parallax measurements 
of Galactic Cepheids~\cite{bened07} gives $75.7 \pm 2.6\kmsmpc$, and using 
the best available sample of LMC Cepheids observed in the infrared 
\cite{perss04} yields $74.4 \pm 2.5\kmsmpc$. Infrared observations are 
important because they reduce the potential error involved in extinction 
corrections. Indeed, the Carnegie Hubble Programme~\cite{freed12} takes 
this further by using mid-IR observations (at 3.6~$\mu$m) of the Benedict 
et al. Galactic Cepheids with measured parallaxes, thus anchoring the 
calibration of the mid-IR \textit{P-L} relation in these objects, and obtaining 
$H_0=74.3 \pm 2.1\kmsmpc$. In the mid-IR, as well as smaller extinction 
corrections, metallicity effects are also generally less. However, arguments
for lower values based on different outlier rejection can give a combined
estimate for the three different calibrations~\cite{efsta13} of
$72.5 \pm 2.5\kmsmpc$.

\subsubsection{SNe Ia systematics}

The calibration of the type Ia supernova distance scale, and hence $H_0$, is
affected by the selection of galaxies used which contain both Cepheids
and historical supernovae. \cite{riess05} make the case for the
exclusion of a number of older supernovae from previous samples with measurements on
photographic plates. Their exclusion, leaving four calibrators with data
judged to be of high quality, has the effect of shrinking the
average distances, and hence raising $H_0$, by a few percent.
\cite{freed01} included six galaxies including SN~1937C, excluded by
\cite{riess05}, but obtained approximately the same value for $H_0$.

Since SNe~Ia occur in galaxies, their brightnesses are likely to be altered
by extinction in the host galaxy. This effect can be assessed and, if necessary,
corrected for, using information about SNe~Ia colours in local SNe. The effect
is found to be smaller than other systematics within the distance ladder
\cite{riess09b}.

Further possible effects include differences in SNe~Ia
luminosities as a function of environment. \cite{wang06} used a
sample of 109 supernovae to determine a possible effect of metallicity
on SNe~Ia luminosity, in the sense that supernovae closer to the centre
of the galaxy (and hence of higher metallicity) are brighter. They
include colour information using the indicator $\Delta C_{12}\equiv
(B-V)_{12\mathrm{\, days}}$, the $B-V$ colour at 12 days after maximum,
as a means of reducing scatter in the relation
between peak luminosity and $\Delta m_{15}$ which forms the traditional
standard candle. Their value of $H_0$ is, however, quite close to the
Key Project value, as they use the four galaxies of~\cite{riess05}
to tie the supernova and Cepheid scales together. This closeness
indicates that the SNe~Ia environment dependence is probably a small
effect compared with the systematics associated with Cepheid
metallicity.

\subsubsection{Other methods of establishing the distance scale}

In some cases, independent distances to
galaxies are available in the form of studies of the tip of the red
giant branch. This phenomenon refers to the fact that metal-poor,
population~II red giant stars have a well-defined cutoff in luminosity
which, in the $I$-band, does not vary much with nuisance parameters such
as stellar population age. Deep imaging can therefore provide an
independent standard candle which can be compared with that of the
Cepheids, and in particular with the metallicity of the Cepheids in
different galaxies. The result~\cite{sakai04} is again that
metal-rich Cepheids are brighter, with a quoted slope of
$-0.24 \pm 0.05 \mathrm{\ mag\ dex}^{-1}$. This agrees with earlier
determinations~\cite{kocha97,kenni03} and is usually adopted when a
global correction is applied.

Several different methods have been proposed to bypass some of the early
rungs of the distance scale and provide direct measurements of distance
to relatively nearby galaxies. Many of these are reviewed in the article
by Olling~\cite{ollin06}.

One of the most promising methods is the use of detached eclipsing
binary stars to determine distances directly~\cite{paczy97}. In nearby
binary stars, where the components can be resolved, the determination of
the angular separation, period and radial velocity amplitude immediately
yields a distance estimate. In more distant eclipsing binaries in other
galaxies, the angular separation cannot be measured directly. However, the
light-curve shapes provide information about the orbital period, the 
ratio of the radius of each star to the orbital separation, and the ratio 
of the stars' luminosities. Radial velocity curves can then be used to
derive the stellar radii directly. If we can obtain a physical handle on the
stellar surface brightness (e.g., by study of the spectral lines) then
this, together with knowledge of the stellar radius and of the observed 
flux received from each star, gives a determination of distance. The
DIRECT project~\cite{bonan06} has used this method to derive
a distance of $964 \pm 54 \mathrm{\ kpc}$ to M33, which is higher than standard
distances of $800\mbox{\,--\,}850 \mathrm{\ kpc}$~\cite{freed91,lee02}. 
It will be interesting to see whether this discrepancy continues
after further investigation.

A somewhat related method, but involving rotations of stars around the 
centre of a distant galaxy, is the method of rotational
parallax~\cite{peter97,ollin00,ollin06}. Here one observes
both the proper motion corresponding to circular rotation, and the
radial velocity, of stars within the galaxy. Accurate measurement of the
proper motion is difficult and will require observations from future
space missions.

\section{The CMB and Cosmological Estimates of the Distance Scale}
\label{sec:CMB}

\subsection{The physics of the anisotropy spectrum and its implications}

The physics of stellar distance calibrators is very complicated, because
it comes from the era in which the Universe has had time to evolve
complicated astrophysics. A large class of alternative approaches to
cosmological parameters in general involve going back to an era where
astrophysics is relatively simple and linear, the epoch of recombination
at which the CMB fluctuations can be studied. Although tests involving
the CMB do not directly determine $H_0$, they provide joint
information about $H_0$ and other cosmological parameters which is
improving at a very rapid rate.

In the Universe's early history, its temperature was high enough to
prohibit the formation of atoms, and the Universe was therefore ionized.
Approximately $4\times 10^5 \mathrm{\ yr}$ after the Big Bang, corresponding to
a redshift $z_\mathrm{rec} \sim 1000$, the temperature
dropped enough to allow the formation of atoms, a point known as
``recombination''. For photons, the consequence of recombination was
that photons no longer scattered from ionized particles but were free to
stream. After recombination, these primordial photons reddened with 
the expansion of the Universe, forming the cosmic microwave background 
(CMB) which we observe today as a black-body radiation background at
$2.73 \mathrm{\ K}$.

In the early Universe, structure existed in the form of 
small density fluctuations ($\delta\rho/\rho\sim0.01$) in the
photon-baryon fluid. The resulting pressure gradients, together with
gravitational restoring forces, drove oscillations, very similar to the
acoustic oscillations commonly known as sound waves. Fluctuations prior
to recombination could propagate at the relativistic ($c/\sqrt{3}$) 
sound speed as the Universe expanded.  At recombination, the structure
was dominated by those oscillation frequencies which had completed a
half-integral number of oscillations within the characteristic size of
the Universe at recombination\epubtkFootnote{This characteristic size is
about 4$\times 10^5$ light years at recombination, corresponding to an angular 
scale of about 1$^{\circ}$ on the sky. The fact that the CMB is homogeneous
on scales much larger than this is an illustration of the ``horizon problem''
discussed in section 1.2, and which inflation may solve.}; this pattern became 
frozen into the photon field which formed the CMB once the photons and baryons 
decoupled and the sound speed dropped. The process is reviewed in much more 
detail in~\cite{hu02}.

The resulting ``acoustic peaks'' dominate the fluctuation spectrum
(see Figure~\ref{figure_04}). Their angular scale is a function of the
size of the Universe at
the time of recombination, and the angular diameter distance between us
and $z_\mathrm{rec}$. Since the angular diameter distance is a function of
cosmological parameters, measurement of the positions of the acoustic
peaks provides a constraint on cosmological parameters. Specifically,
the more closed the spatial geometry of the Universe, the smaller the
angular diameter distance for a given redshift, and the larger the
characteristic scale of the acoustic peaks. The measurement of the peak
position has become a strong constraint in successive observations 
(in particular BOOMERanG~\cite{deber00}, MAXIMA ~\cite{hanan00},
and WMAP, reported in~\cite{sperg03} and~\cite{sperg06}) and corresponds to an
approximately spatially flat Universe in which
$\Omega_\mathrm{m}+\Omega_{\Lambda}\simeq1$.

\epubtkImage{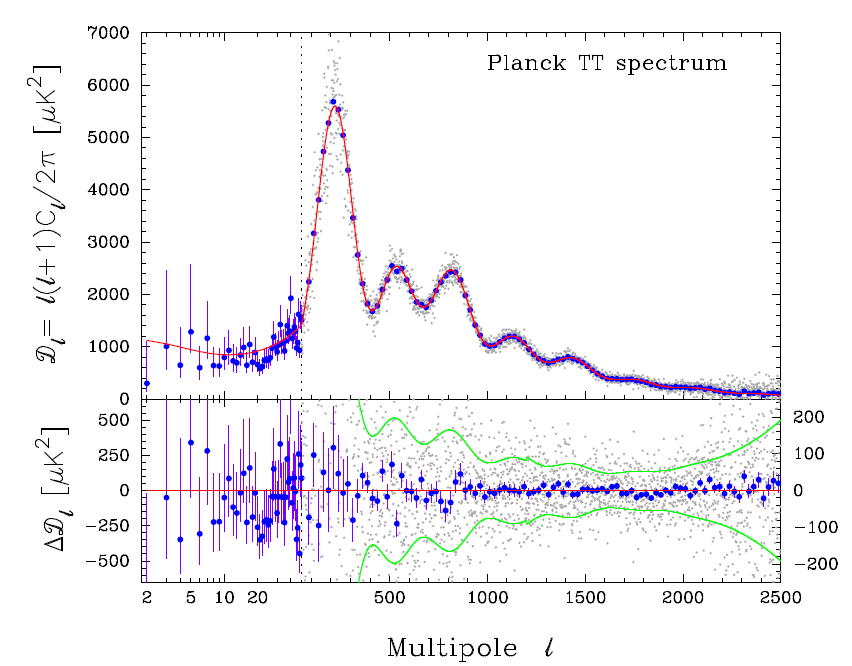}{%
  \begin{figure}[htbp]
    \centerline{\includegraphics[scale=0.6]{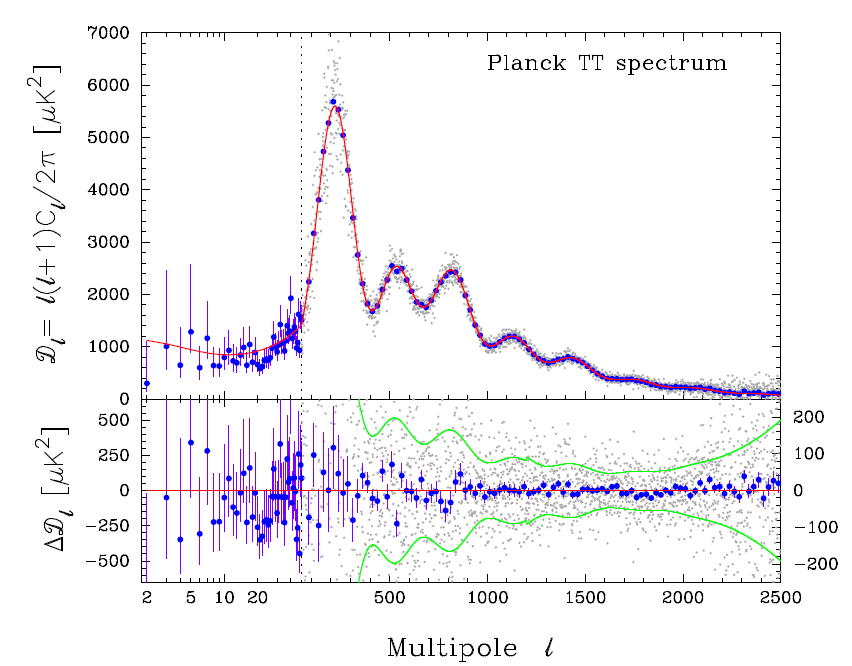}}
    \caption{Diagram of the CMB anisotropies, plotted as strength 
      against spatial frequency, from the 2013 Planck data~\cite{planc13}. The
      measured points are shown together with best-fit models. Note the
      acoustic peaks, the largest of which corresponds to an angular 
      scale of about half a degree.}
    \label{figure_04}
\end{figure}}

But the global geometry of the Universe is not the only property which
can be deduced from the fluctuation spectrum.\epubtkFootnote{See 
\url{http://background.uchicago.edu/~whu/intermediate/intermediate.html}
for a much longer exposition and tutorial on all these areas.} The peaks are also
sensitive to the density of baryons, of total (baryonic + dark) matter,
and of vacuum energy (energy associated with the cosmological constant).
These densities scale with the square of the Hubble parameter times the
corresponding dimensionless densities (see
Equation~(\ref{equation_05})) and measurement of the acoustic peaks
therefore provides information on the Hubble constant, degenerate with
other parameters, principally $\Omega_{\mathrm{m}}$ and $\Omega_{\Lambda}$. The second peak
strongly constrains the baryon density, $\Omega_{\mathrm{b}} H_0^2$, 
and the third peak is sensitive to the total matter density in the 
form $\Omega_{\mathrm{m}} H_0^2$.


\subsection{Degeneracies and implications for \textit{H}\sub{0}}

Although the CMB observations provide significant information about cosmological
parameters, the available data constrain combinations of $H_0$ with other parameters,
and either assumptions or other data must be provided in order to derive the Hubble 
constant. One possible assumption is that the Universe is exactly flat (i.e., $\Omega_k=0$) 
and the primordial fluctuations have a power law spectrum. In this case measurements of 
the CMB fluctuation spectrum with the Wilkinson Anisotropy Probe (WMAP) satellite 
\cite{sperg03,sperg06} and more recently with the Planck satellite~\cite{planc13}, allow 
$H_0$ to be derived. This is because measuring $\Omega_{\mathrm{m}}h^2$ produces a locus in the
$\Omega_{\mathrm{m}}:\Omega_{\Lambda}$ plane 
which is different from the $\Omega_{\mathrm{m}}+\Omega_{\Lambda}=1$
locus of the flat Universe, and although the tilt of these two lines is not very different,
an accurate CMB measurement can localise the intersection enough to give $\Omega_{\mathrm{m}}$ and
$h$ separately. The value of $H_0 = 73 \pm 3\kmsmpc$ was derived in this way by WMAP 
\cite{sperg06} and, using the more accurate spectrum provided by Planck, as 
$H_0=67.3 \pm 1.2\kmsmpc$~\cite{planc13}. In this case, other cosmological parameters can 
be determined to two and in some cases three significant figures,\epubtkFootnote{See Table 2 
of~\cite{planc13} for a full list.} Figure~\ref{figure_05} shows this in another way,
in terms of $H_0$ as a function of $\Omega_m$ in a flat universe ($\Omega_k=0$ in 
Equation~\ref{equation_07C}).

\epubtkImage{}{%
  \begin{figure}[htbp]
    \centerline{\includegraphics[width=10cm]{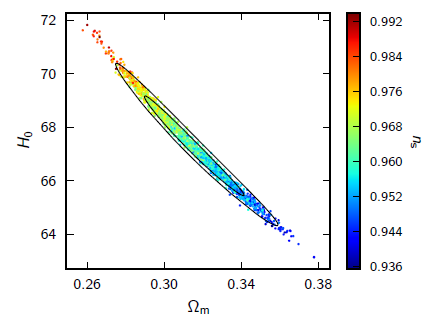}}
\caption{The allowed range of the parameters $H_0$, $\Omega_\mathrm{m}$, from the 2013 
Planck data, is shown as a series of points (reproduced from~\cite{planc13}). A flat
Universe is assumed, together with information from the Planck fluctuation temperature
spectrum, CMB lensing information from Planck, and WMAP polarization observations. The
colour coding reflects different values of $n_s$, the spectral index of scalar perturbations
as a function of spatial scale at early times.
}
    \label{figure_05}
\end{figure}}

If we do not assume the universe to be exactly flat, as is done in Figure~\ref{figure_05}, 
then we obtain a degeneracy with $H_0$ in the sense that decreasing $H_0$ increases the total 
density of the universe (approximately by 0.1 in units of the critical density for a
20\kmsmpc\ decrease in $H_0$). CMB data by themselves, 
without any further assumptions or extra data, do not supply a significant 
constraint on $H_0$ compared to those which are obtainable by other methods.
Other observing programmes are, however, available which result in constraints on
the Hubble constant together with other parameters, notably $\Omega_{\mathrm{m}}$, $\Omega_{\Lambda}$ 
and $w$ (the dark energy equation of state parameter defined in section 1.2); we can either 
regard $w$ as a constant or allow a variation with redshift. We
sketch these briefly here; a full review of all programmes addressing cosmic acceleration
can be found in the review by Weinberg et al.~\cite{weinb13}.

The first such supplementary programme is the study of type Ia supernovae, which as we 
have seen function as standard candles (or at least easily calibratable candles). They 
therefore determine the luminosity distance $D_L$. Studies of SNe~Ia were the first 
indication that $D_L$ varies with $z$ in such a way that an acceleration term, 
corresponding to a non-zero $\Omega_{\Lambda}$ is required~\cite{perlm97,perlm98,riess98}, 
a discovery which won the 2011 Nobel Prize in physics. This determination of luminosity 
distance gives constraints in the $\Omega_{\mathrm{m}}:\Omega_{\Lambda}$ plane which are more or 
less orthogonal to the CMB constraints. Currently, the most complete samples of distant
SNe come from SDSS surveys at low redshift ($z < 0.4$)~\cite{friem08,sako08,holtz08,kessl09}, 
the ESSENCE survey at moderate redshift ($0.1 < z < 0.78$)~\cite{mikna07,woodv07}, the
SNLS surveys at $z < 1$~\cite{conle11} and high-redshift ($z > 0.6$) HST surveys 
\cite{riess07,dawso09,suzuk12}. In the future, surveys in the infrared should be capable
of extending the redshift range further~\cite{rodne12}.

The second important programme is the measurement of structure at more
recent epochs than the epoch of recombination using the characteristic
length scale frozen into the structure of matter at recombination (section
4.1). This is manifested in the real Universe by an expected
preferred correlation length of $\sim$~100~Mpc between
observed baryon structures, otherwise known as galaxies. These baryon
acoustic oscillations (BAOs) measure a standard rod, and constrain the
distance measure $D_V \equiv (cz(1+z)^2D_A^2H(z)^{-1})^{1/3}$
(e.g.~\cite{eisen05}). The largest sample available for such studies
comes from luminous red galaxies (LRGs) in the Sloan Digital Sky
Survey~\cite{york00}. The expected signal was first
found~\cite{eisen05} in the form of an increased power in the
cross-correlation between galaxies at separations of about 100~Mpc, and
corresponds to an effective measurement of angular diameter distance
to a redshift $z\sim0.35$. Since then, this characteristic distance
has been found in other samples at different redshifts, 6dFGS at
$z\simeq 0.1$~\cite{beutl11}, further SDSS work at
$z=0.37$~\cite{padma12} and by the BOSS and WiggleZ collaborations at
$z\simeq 0.6$~\cite{blake11,ander12}. It has also been observed in studies 
of the Ly$\alpha$ forest~\cite{busca13,slosa13,delub14}. In principle, provided
the data are good enough, the BAO can be studied separately in the radial and 
transverse directions, giving separate constraints on $D_A$ and 
$H(z)$~\cite{sanch13,bull14} and hence more straightforward and accurate
cosmology.

There are a number of other programmes which constrain combinations of cosmological parameters
which can break degeneracies involving $H_0$. Weak lensing observations have progressed very
substantially over the last decade, after a large programme of quantifying and reducing 
systematic errors; these observations consist of measuring shapes of large numbers of 
galaxies in order to extract the small shearing signal produced by matter along the line
of sight. The quantity directly probed by such observations is a combination of 
$\Omega_{\mathrm{m}}$ and $\sigma_8$, the rms density fluctuation at a scale of $8h^{-1}$~Mpc. 
State-of-the-art
surveys include the CFHT survey~\cite{heyma12,kilbi13} and SDSS-based surveys
\cite{lin12}. Structure can also be mapped using Lyman-$\alpha$ forest observations. 
The spectra of distant quasars have deep absorption lines corresponding to absorbing matter
along the line of sight. The distribution of these lines measures clustering of matter on 
small scales and thus carries cosmological information (e.g.~\cite{tytle04,mcdon05}). 
Clustering on small scales~\cite{tegma04b} can be mapped, and the matter power spectrum 
can be measured, using large samples of galaxies, giving constraints on combinations of $H_0$, 
$\Omega_{\mathrm{m}}$ and $\sigma_8$.


\subsubsection{Combined constraints}

As already mentioned, Planck observations of the CMB alone are capable
of supplying a good constraint on $H_0$, given three assumptions: the
curvature of the Universe, $\Omega_k$, is zero, that dark energy is a
cosmological constant ($w=-1$) and that it is independent of redshift
($w\neq w(z)$). In general, every other accurate measurement of a
combination of cosmological parameters allows one to relax one of the
assumptions. For example, if we admit the BAO data together with the
CMB, we can allow $\Omega_k$ to be a free
parameter~\cite{tegma06,ander12,planc13}. Using earlier WMAP data for
the CMB, $H_0$ is derived to be $69.3 \pm
1.6\kmsmpc$~\cite{mehta12,ander12} which does not change significantly
using Planck data ($68.4 \pm 1.0\kmsmpc$~\cite{planc13}); the
curvature in each case is tightly constrained (to $<0.01$) and
consistent with zero. If we introduce supernova data instead of BAO
data, we can obtain $w$ provided that
$\Omega_k=0$~\cite{weinb13,planc13} and this is found to be consistent
with $w=-1$ within errors of about 0.1\,--\,0.2~\cite{planc13}.

If we wish to proceed further, we need to introduce additional data to
get tight constraints on $H_0$. The obvious option is to use both BAO
and SNe data together with the CMB, which results in $H_0 = 68.7 \pm
1.9\kmsmpc$~\cite{blake11} and $69.6 \pm 1.7$ (see Table~4
of~\cite{ander12}) using the WMAP CMB constraints. Such analyses
continue to give low errors on $H_0$ even allowing for a varying $w$
in a non-flat universe, although they do use the results from three
separate probes to achieve this. Alternatively, extrapolation of the
BAO results to $z = 0$ give $H_0$
directly~\cite{eisen98,beutl11,weinb13} because the BAO measures a
standard ruler, and the lower the redshift, the purer the standard
ruler's dependence on the Hubble constant becomes, independent of
other elements in the definition of Hubble parameter such as
$\Omega_k$ and $w$. The lowest-redshift BAO measurement is that of the
6dF, which suggests $H_0 = 67.0 \pm 3.2\kmsmpc$~\cite{beutl11}.


\newpage


\section{Conclusion}
\label{sec:conclusion}

Progress over the last few years in determining the Hubble constant to
increasing accuracy has been encouraging and rapid. For the first
time, in the form of megamaser studies, there is a one-step method
available which does not have serious systematics. Simultaneously,
gravitational lens time delays, also a one-step method but with a
historical problem with systematics due to the mass model, has also
made progress due to a combination of better simulations of the
environment of the lens galaxies and better use of information which
helps to ease the mass degeneracy. The classical Cepheid method has
also yielded greatly improved control of systematics, mainly by moving
to calibrations based on NGC~4258 and Galactic Cepheids which are much
less sensitive to metallicity effects.

Identification of the current ``headline'' best $H_0$ distance
determinations, by methods involving astrophysical objects, is a
somewhat subjective business. However, most such lists are likely to
include the following, including the likely update paths:

\begin{itemize}
\item Megamasers: $68.0 \pm 4.8\kmsmpc$~\cite{braat13}. Further
  progress will be made by identification and monitoring of additional
  targets, since the systematics are likely to be well controlled
  using this method.
\item Gravitational lenses: $73.1^{+2.4}_{-3.6}\kmsmpc$~\cite{suyu13a}
  (best determination with systematics controlled, 2 lenses), $69 \pm
  6/4\kmsmpc$~\cite{seren14} (18 lenses, but errors from range of
  free-form modelling). Progress is likely by careful control of
  systematics to do with the lens mass model and the surroundings in
  further objects; a programme 
(H0LiCOW~\cite{suyu13}) is beginning with precisely this objective.
\item Cepheid studies: $72.0 \pm 3.0\kmsmpc$~\cite{riess11} with
  corrected NGC~4258 distance from~\cite{humph13}; $75.7 \pm
  2.6\kmsmpc$ (parallax of Galactic Cepheids) and $74.3 \pm
  2.1\kmsmpc$ (mid-IR observations)~\cite{freed12}. The Carnegie
  Cepheid programme is continuing IR observations which should
  significantly reduce systematics of the method.
\end{itemize}

In parallel with these developments, the Planck satellite has given us
much improved constraints on $H_0$ in combination with other
cosmological parameters. The headline $H_0$ determinations are all
from Planck in combination with other information, and are:

\begin{itemize}
\item For a flat-by-fiat Universe, $H_0=67.3 \pm
  1.2\kmsmpc$~\cite{planc13} from Planck.
\item For a Universe free to curve, $H_0=68.4 \pm
  1.0\kmsmpc$~\cite{planc13} using Planck together with BAO data.
\item Local BAO measurements: $H_0=67.0 \pm 3.2\kmsmpc$~\cite{beutl11}
  using only the well-determined $\Omega_{\mathrm{m}}h^2$ from the
  CMB, but independent of other
  cosmology. \cite{beutl11,blake11,ander12,padma12}
\end{itemize}

There is thus a mild tension
between some (but not all) of the astrophysical measurements and the
cosmological inferences. There are several ways of looking at
this. The first is that a 2.5-$\sigma$ discrepancy is nothing to be
afraid of, and indeed is a relief after some of the clumped
distributions of published measurements in the past. The second is that
one or more methods are now systematics-limited; in other words, the
subject is limited by accuracy rather than precision, and that 
careful attention to underestimated systematics will cause the values
to converge in the next few years. Third, it is possible that new physics is
involved beyond the variation of the dark energy index $w$. This new
physics could, for example, involve the number of relativistic degrees
of freedom being greater than the standard value of 3.05,
corresponding to three active neutrino contributions~\cite{planc13}; or
a scenario in which we are living in a local bubble with a different
$H_0$~\cite{marra13}. Most instincts would dictate taking these
possibilities in this order, unless all of the high-quality
astrophysical $H_0$ values differed from the cosmological ones.

The argument can be turned around, by observing that independent
determinations of $H_0$ can be fed in as constraints to unlock a
series of accurate measurements of other cosmological parameters such
as $w$. This point has been made a number of times, in particular by
Hu~\cite{hu05}, Linder~\cite{linde11} and Suyu et al.~\cite{suyu13b};
the dark energy figure of merit, which measures the $P-\rho$
dependence of dark energy and its redshift evolution, can be be
improved by large factors using such independent measurements. Such
measurements are usually extremely cheap in observing time (and
financially) compared to other dark energy programmes. They will,
however, require 1\% determinations of $H_0$, given the current state
of play in cosmology. This is not impossible, and should be reachable
with care quite soon.

\newpage

\section{Acknowledgements}

I thank Adam Bolton and a second anonymous referee for comments on this
version of the paper, and Ian Browne, Nick Rattenbury and Sir Francis
Graham-Smith for discussions and comments on the original version.


\bibliography{refs}

\end{document}